\documentclass[pra,twocolumn,showpacs,eqsecnum]{revtex4}

\usepackage{graphicx}
\usepackage[urlcolor=blue, hyperindex, colorlinks, bookmarks=true,linkcolor=black,citecolor=black]{hyperref}

\begin{document}

\newcommand{\sch}{Schr\"odinger }
\newcommand{\schs}{Schr\"odinger's }
\newcommand{\beq}{\begin{equation}}
\newcommand{\eeq}{\end{equation}}
\newcommand{\bqa}{\begin{eqnarray}}
\newcommand{\eqa}{\end{eqnarray}}
\newcommand{\nn}{\nonumber}
\newcommand{\nl}{\nn \\ &&}
\newcommand{\dg}{^\dagger}
\newcommand{\rt}[1]{\sqrt{#1}\,}
\newcommand{\smallfrac}[2]{\mbox{$\frac{#1}{#2}$}}
\newcommand{\bra}[1]{\langle{#1}|}
\newcommand{\ket}[1]{|{#1}\rangle}
\newcommand{\ito}{It\^o }
\newcommand{\str}{Stratonovich }
\newcommand{\bl}{{\Bigl(}}
\newcommand{\br}{{\Bigr)}}

\newcommand{\erf}[1]{Eq.~(\ref{#1})}
\newcommand{\erfs}[2]{Eqs.~(\ref{#1}) and (\ref{#2})}
\newcommand{\erft}[2]{Eqs.~(\ref{#1}) -- (\ref{#2})}
\newcommand{\lfrac}[2]{({#1})/{#2}}

\newcommand{\qk}{\ket{\{q_{k}\}}}
\newcommand{\qb}{\bra{\{q_{k}\}}}
\newcommand{\psiqk}[1]{\ket{{\psi}_{\{q_{k}\}}({#1})}}
\newcommand{\psiqb}[1]{\bra{{\psi}_{\{q_{k}\}}({#1})}}
\newcommand{\lpsiqk}[1]{\ket{\tilde{\psi}_{\{q_{k}\}}({#1})}}
\newcommand{\lpsiqb}[1]{\bra{\tilde{\psi}_{\{q_{k}\}}({#1})}}
\newcommand{\ak}{\ket{\{a_{k}\}}}
\newcommand{\ab}{\bra{\{a_{k}\}}}
\newcommand{\psiak}[1]{\ket{{\psi}_{\{a_{k}\}}({#1})}}
\newcommand{\psiab}[1]{\bra{{\psi}_{\{a_{k}\}}({#1})}}
\newcommand{\lpsiak}[1]{\ket{\tilde{\psi}_{\{a_{k}\}}({#1})}}
\newcommand{\lpsiab}[1]{\bra{\tilde{\psi}_{\{a_{k}\}}({#1})}}
\newcommand{\psizk}[1]{\ket{\psi_{z}({#1})}}
\newcommand{\psizb}[1]{\bra{\psi_{z}({#1})}}
\newcommand{\lpsizk}[1]{\ket{\tilde{\psi}_{z}({#1})}}
\newcommand{\lpsizb}[1]{\bra{\tilde{\psi}_{z}({#1})}}
\newcommand{\psiIk}[1]{\ket{\psi_{I}({#1})}}
\newcommand{\psiIb}[1]{\bra{\psi_{I}({#1})}}
\newcommand{\lpsiIk}[1]{\ket{\tilde{\psi}_{I}({#1})}}
\newcommand{\lpsiIb}[1]{\bra{\tilde{\psi}_{I}({#1})}}

\title{Non-Markovian stochastic \sch equations: Generalization to
real-valued noise using quantum measurement theory}
\date{\today}
\author{Jay Gambetta}
\affiliation{Centre for Quantum Dynamics, School of Science,
Griffith University, Brisbane 4111, Australia}
\author{H. M. Wiseman} \email{h.wiseman@gu.edu.au}
\affiliation{Centre for Quantum Dynamics, School of Science,
Griffith University, Brisbane 4111, Australia}

\begin{abstract}
Do stochastic \sch equations, also known as unravelings, have a physical interpretation? In the Markovian limit, where the system {\em on average} obeys a master equation, the answer is yes. Markovian stochastic \sch equations generate quantum trajectories for the system state conditioned on continuously monitoring the bath. For a given master equation, there are many different unravelings, corresponding to different sorts of measurement on the bath. In this paper we address the non-Markovian case, and in particular the sort of stochastic \sch equation introduced by Strunz, Di\' osi, and Gisin [Phys. Rev. Lett. {\bf 82}, 1801 (1999)]. Using a quantum measurement theory approach, we rederive their unraveling which involves complex-valued Gaussian noise. We also derive an unraveling involving real-valued Gaussian noise. We show that in the Markovian limit, these two unravelings correspond to heterodyne and homodyne detection respectively. Although we use quantum measurement theory to define these unravelings, we conclude that the stochastic evolution of the system state is not a true quantum trajectory, as the identity of the state through time is a fiction.
\end{abstract}

\maketitle

\section{Introduction}
In nature, a quantum system is most likely found in an entanglement with at
least one other
quantum system. An example of this is a two level atom (TLA) immersed in an
environment of
harmonic oscillators (the electromagnetic field). This type of quantum
system, a small
system interacting with a larger system (the bath) is called an open
quantum system
\cite{Car93}. The system-bath interaction causes the two systems to
entangle, resulting in
a combined state $\ket{\Psi(t)}$ whose evolution can be theoretically
determined by the
\sch equation. However, due to the many degrees of freedom of the bath,
this is generally
impractical and it is best to describe the system (TLA) by the reduced
state $\rho_{\rm
red}(t)$. The evolution of $\rho_{\rm red}(t)$ is found by averaging the
outer product of
the \sch equation over all the possible bath states,
\begin{equation}\label{ReducedState}
 \rho_{\rm red}(t)={\rm Tr}_{\rm field}[\ket{\Psi(t)}\bra{\Psi(t)}].
\end{equation}

Under the Born-Markov approximations \cite{Gar91} it is possible
to obtain a closed equation for $\rho_{\rm red}(t)$. For
mathematical consistency, this should be of the Lindblad form
\cite{Lin76}. If there is a single Lindblad operator $\hat{L}$
(such as the lowering operator for the system) then this is an
equation of the form
\begin{equation}\label{MarkovReducedState}
\dot\rho_{\rm red} (t) =-i[\hat{H},\rho_{\rm
red}(t)]+\gamma {\cal D}[\hat{L}]\rho_{\rm red} (t)
,
\end{equation} where $\hat{H}$ is the Hamiltonian and
\begin{equation}\label{DOperator}
{\cal D}[\hat{L}]\rho_{\rm red}=\hat{L}\rho_{\rm
red}\hat{L}\dg-\smallfrac12 \hat{L}\dg \hat{L}\rho_{\rm red}-\smallfrac12
\rho_{\rm red}\hat{L}\dg\hat{L}.
\end{equation}
However this is only an approximation, in the non-Markovian situation in general one can
not solve $\rho_{\rm red}(t)$ or $\ket{\Psi(t)}$ analytically, so $\rho_{\rm red}(t)$ is
difficult to determine.

A breakthrough in solving this problem was achieved with the
development of non-Markovian stochastic \sch~ equations (SSEs).
These stochastic differential equations for a state vector were
first introduced for Markovian open quantum systems in
mathematical physics
\cite{Dio88,Dio88b,Bel88,BelSta92,Bar90,Bar93,GisPer92,GisPer92b,GisPer93}
and then independently in quantum optics
\cite{Car93,DalCasMol92,GarParZol92}. This approach has
subsequently been generalized to deal with non-Markovian systems
\cite{Ima94,Str96,JacCol00,Cre00,Dio96}. In this paper we will
follow the approach of Di\'osi, Strunz and Gisin (DSG)
\cite{Dio96,DioStr97,DioGisStr98,StrDioGis99}. In their approach,
the system state vector $\psizk{t}$ \footnote{A subscript $z$ on a
parameter means it is a functional of $z(t)$} depends upon some
(not necessarily white) noise $z(t)$, which is drawn from some
probability distribution. The SSE has the property that when the
outer product of $\psizk{t}$ is averaged over all the possible
$z(t)$ one obtains $\rho_{\rm red}$(t). That is,
\begin{equation}\label{Ensemble}
\rho_{\rm red}(t)={\rm E}[\psizk{t}\psizb{t}],
\end{equation}
where E[...] denotes an ensemble average over all possible $z(t)$'s.

In cases where an exact non-Markovian SSE can be derived, it is also possible to find an
exact solution for $\rho_{\rm red}(t)$
 by other means. A key advantage of non-Markovian SSEs lies in the
 cases where no exact solution is possible. In this case approximations
must be made in either approach. The advantage of the SSE approach is that the ensemble
average $\rho_{\rm red}(t)$ is, by construction, guaranteed to be a positive operator.
This fundamental property of a state matrix is not guaranteed by other approximate
equations for $\rho_{\rm red}(t)$. This is true even in in the Markov limit; quantum
Brownian motion is a case in point \cite{StrDioGisYu99}. The other advantage of the SSE
approach in general is that it allows the evolution of large systems to be simulated
numerically. This was the original motivation for their introduction in quantum optics
\cite{DalCasMol92,GarParZol92}.

Leaving aside the potential usefulness of SSEs,
one may ask the question: is there a physical interpretation for the
solution of an SSE, or is it simply a
numerical tool for finding $\rho_{\rm red} (t)$? In the Markovian
limit, that is when the master equation has the form \erf{MarkovReducedState},
 the answer is yes. The solution to the SSE, termed by Carmichael is a
 {\em quantum trajectory} \cite{Car93}, it can be interpreted as the state of
 the system conditioned on the measurement
results obtained by continuously monitoring the bath \cite{WisMil93}. For the Markovian
case, different sorts of SSEs exist. They may involve jumps or diffusion, and are termed
different {\em unravelings} of the master equation \cite{Car93}. These different
unravelings correspond to different detection schemes, such as photon counting
\cite{Car93,DalCasMol92,GarParZol92}, homodyne \cite{Car93,WisMil93,WisMil93c}, and
heterodyne \cite{WisMil93c} detection. Other generalizations
\cite{Wis96,JacColWal99,GamWis01,WarWisMab01} have also been investigated.

In this paper we will investigate the question of physical
interpretation of {\em non-Markovian} diffusive SSEs
of Diosi, Strunz and Gisin (DSG)
\cite{Str96,Dio96,DioStr97,DioGisStr98,StrDioGis99}. We will show
that quantum measurement theory (QMT) does give meaning to the
$\psizk{t}$ at any particular time, $t$. However, the linking of
the state $\psizk{t}$ at different times to make a trajectory
appears to be a convenient fiction. We also show that the theory
of DSG  can
be generalized by considering different sorts of measurements
(unravelings) on the bath. We use our approach to define two
different unravelings. The first results in DSG SSEs, with
complex-valued noise $z(t)$. In the Markovian limit
this unraveling corresponds to heterodyne detection. The second,
which can only be defined for some system-bath couplings, has
real-valued noise and has homodyne detection as its Markovian limit.

\section{ System dynamics and Quantum Measurement Theory.}

\subsection{\sch Equation for the combined system}\label{Dynamics}

With $\hbar=1$, a system interacting with a reservoir of harmonic
oscillators has the
total Hamiltonian
\begin{equation} \label{HamiltonianTotal}
\hat{H}_{\rm tot}=\hat{H}_{0}+\hat{H}+\hat{H}_{\rm bath}+\hat{V}.
\end{equation}
Here the system Hamiltonian has been split into $\hat{H}_{0}$ (the
action of which is described later) and
$\hat{H}$ (the remainder). The Hamiltonian for the bath is
\begin{equation}\label{HamiltonianBath}
\hat{H}_{\rm bath}=\sum_{k}\omega_{k}\hat{a}_{k}\dg \hat{a}_{k},
\end{equation} where $k$ labels the modes of the bath,
$\hat{a}_{k}$ and $\omega_{k}$ are the lowering operator and angular
frequency of the
$k^{\rm th}$ mode respectively.
We assume the
interaction Hamiltonian to have the form
\begin{equation}\label{HamiltonianInteraction}
\hat{V}=i(\hat{L}\hat{b}\dg-\hat{b}\hat{L}\dg),
\end{equation}
where $\hat{L}$ is a system operator and
where we have define the bath lowering operators $\hat{b}$
as $\hat{b}=\sum_{k}g_{k}\hat{a}_{k}$. That is, the coupling
amplitude of the $k^{\rm th}$ mode to the system is $g_{k}$.

The \sch equation for the combined state is
\begin{equation}\label{SchEquation}
d_{t}\ket{{\Psi}(t)}=-{i}\hat{H}_{\rm tot}\ket{\Psi(t)},
\end{equation} which can equivalently be written as
\begin{equation}\label{UnitarySch}
  \ket{\Psi(t)}=U(t,0)\ket{\Psi(0)},
\end{equation} where $U(t,0)$ is called the unitary evolution operator.
Defining a unitary evolution operator for the `free' system and bath as
\begin{equation}\label{UnitaryFree}
U_{0}(t,0)=e^{-i(\hat{H}_{0}+\hat{H}_{\rm bath})(t-0)}.
\end{equation}
We can write $U(t,0)$ as  $U(t,0)=U_{0}(t,0)U_{\rm
int}(t,0)$, where $U_{\rm int}(t,0)$ is the unitary evolution
operator that describes the total evolution with the free dynamics
removed.

We can then define an interaction picture state as
\begin{equation} \label{InteractionState}
\ket{\Psi_{\rm int}(t)}=U_{\rm int}(t,0)\ket{\Psi(0)},
\end{equation}
which obeys
\begin{equation} \label{IntSchEquation}
d_t\ket{{\Psi}_{\rm int}(t)} =-{i}\bl \hat{H}_{\rm int}(t)+
\hat{V}_{\rm int}(t) \br \ket{\Psi_{\rm int}(t)}.
\end{equation}  The Hamiltonians in the interaction picture are
\begin{eqnarray} \label{HamiltonianExtraInt}
\hat{H}_{\rm int}(t)&=& U_{0}\dg(t,0)\hat{H}U_{0}(t,0),
\end{eqnarray} and
\begin{eqnarray} \label{HamiltonianInteractionInt}
\hat{V}_{\rm int}(t)&=& i\bl \hat{b}_{\rm int}\dg(t) \hat{L}_{\rm int}(t)
-\hat{b}_{\rm int}(t)\hat{L}_{\rm int}\dg(t) \br,
\end{eqnarray} where
\begin{eqnarray} \label{BathOperatorInt}
\hat{b}_{\rm int}(t)&=&\sum_{k}g_{k}\hat{a}_{k}e^{-i\omega_{k} t}, \\
 \label{SystemOperatorInt}
\hat{L}_{\rm int}(t)&=&\hat{L}e^{-i\omega_{0} t}.
\end{eqnarray}
Here we have finally restricted $\hat{H}_{0}$ to be such that
$\hat{L}$ in the interaction picture simply rotates in the complex
plane as indicated in \erf{SystemOperatorInt}.
The interaction picture can be viewed as moving the time dependencies due
to the free bath
and system dynamics from the state to the operators. Unless otherwise
stated the rest of
this paper will be in the interaction picture and thus we will drop the
subscripts `int'.

\subsection{QMT and Conditional System States}\label{QMT}

In open quantum systems a measurement is always perform on the bath. Due to the
entanglement between the bath and the system the measurement on the bath results in an
indirect measurement of the system \cite{BraKha92}. The state of the system after the
measurement is dependent on the results of the measurement, so we call this a conditional
system state. To mathematically describe this (for a more detailed description see
\cite{Wis96,BraKha92,Kra83}) we define $\qk$ as the arbitrary basis the measurement is
performed in. Note that $\qk$ does not necessarily have to be normalized. For our purposes
we restrict $\qk$ to be a state in the interaction picture with no time dependence (it
will be $U_{0}\dg(t,0)\qk$ in the \sch picture). A typically example of this is a
coherent bath state. This is the state (in the interaction picture) the bath (harmonic
oscillators) has when driven by a classical current \cite{ScuZub97}.

In the basis $\qk$ we can define a probability-operator-measure (POM) element, or effect,
as
\begin{equation}\label{Effect}
\hat{F}_{\{q_{k}\}}=\qk\qb.
\end{equation}
Here the  subscript $\{q_{k}\}$ is the result of the  measurement. The
effect is
important as it allows one to calculate the probability density of results
$\{q_{k}\}$
\begin{equation} \label{Probability}
{P}(\{q_{k}\},t)=\bra{\Psi(t)}\hat{F}_{\{q_{k}\}}\ket{\Psi(t)}.
\end{equation}
If one were only interested in obtaining probabilities the effect would be all one would
need. However, since we are interested in the state of the system after the measurement,
we need to define a set of measurement operators. The constraint the measurement operators
must obey is $\hat{F}_{\{q_{k}\}}=\hat{M}_{\{q_{k}\}}\dg\hat{M}_{\{q_{k}\}}$. For example,
we can decompose the measurement operators as
\begin{equation}\label{MeasuremntOperator}
\hat{M}_{\{q_{k}\}}=\ket{\{n_{k}\}}\qb,
\end{equation} where $\{n_{k}\}$ is arbitrary, and is the
state the bath is left in after the measurement. Since in most detection
situations a
measurement results in annihilating the detected field the most natural
choice for $\{n_{k}\}$
is the vacuum state $\{0_{k}\}$.

In QMT the combined state after a measurement at time $t$, which
yielded results ${\{q_{k}\}}$ is \cite{BraKha92,Kra83}
\begin{equation}\label{MeasurementState}
\ket{\Psi_{\{q_{k}\}}(t)}=\frac{\hat{M}_{\{q_{k}\}}\ket{\Psi(t)}}{\rt{{P}(\{q_{k
}\},t)}}.
\end{equation} Using equation (\ref{MeasuremntOperator}), with $n_{k}=0$
for all $k$, the
combined state after the measurement is,
$\ket{\Psi_{\{q_{k}\}}(t)}=\ket{\{0_{k}\}}\psiqk{t}$, where
\begin{equation}\label{MeasurementSystemState}
\psiqk{t}=\frac{\qb\Psi(t)\rangle}{\rt{{P}(\{q_{k}\},t)}}.
\end{equation}
Equation (\ref{MeasurementSystemState}) is the conditional system state and
we see here directly how the entanglement between the bath and the system
results in the
system state collapsing upon measurement of the bath. One of the properties
of this
conditional system state is that $\rho_{\rm red}(t)$ (equation
(\ref{ReducedState})) can
be written as
\begin{eqnarray} \label{EnsembleConditionalState}
\rho_{\rm red}(t)&=& \int \qb\Psi(t)\rangle\langle\Psi(t)\qk d\{q_{k}\}\nn
\\&=&  \int {P
}(\{q_k\},t)\psiqk{t}\psiqb{t} d\{q_{k}\}\nn \\&=& {\rm E}[\psiqk{t}\psiqb{t}],
\end{eqnarray} where E denote an average over the distribution ${P
}(\{q_k\},t)$. From equation (\ref{Ensemble}) we see that the conditional
state satisfies
the same requirements as a solution of a SSE. This suggests that the time
derivative of
equation (\ref{MeasurementSystemState}), if it could be written in terms
of $\psiqk{t}$, could be interpreted as a SSE. One problem in
determining the time-derivative is that \erf{MeasurementSystemState}
involves the probability ${P}(\{q_{k}\},t)$, which requires knowing
$\ket{\Psi(t)}$,
and, as mention earlier, this in general is indeterminable.
However, this  problem may be overcome using linear quantum measurement
theory (LQMT).

LQMT uses the same principles as QMT except we use an ostensible distribution
($\Lambda(\{q_{k}\})$) in place of the actual probability \cite{Wis96,GoeGra94}. As its
name suggests, the ostensible probability distribution need bear no relation to the actual
probability distribution. However, it must be a proper probability distribution
(non-negative, and integrating to unity), and must be non-zero wherever the actual
distribution is non-zero. Using the ostensible probability distribution, the conditioned
system state is
\begin{equation}\label{LMeasurementSystemState}
\lpsiqk{t}=\frac{\qb\Psi(t)\rangle}{\rt{{ \Lambda}(\{q_{k}\})}}.
\end{equation}
We will call it the linear conditioned system state, because it depends linearly on the
pre-measurement state $\ket{\Psi(t)}$, unlike \erf{MeasurementSystemState}. Since
$\Lambda(\{q_{k}\})$ is not equal to the actual probability, $\lpsiqk{t}$ will not be
normalized and to signify this we use a tilde above the state.  Note that this notation,
following our earlier convention \cite{Wis96,GamWis01}, is the reverse of that used by DSG
\cite{DioGisStr98}. Because it is unnormalized, the linear conditioned system state does
not have a clear physical interpretation. However, it still is useful as it is easier to
calculate (involving only linear equations), and we can write
\begin{eqnarray}\label{LEnsembleConditionalState}
\rho_{\rm red}(t)&=&\int \qb\Psi(t)\rangle\langle\Psi(t)\qk d\{q_{k}\}\nn
\\&=& \int
{\Lambda }(\{q_k\})\lpsiqk{t}\lpsiqb{t} d\{q_{k}\}\nn \\&=&  {\rm
\tilde{E}}[\lpsiqk{t}\lpsiqb{t}], \label{deftildeE}
\end{eqnarray} where ${\rm \tilde{E}}$ denote an average using the
ostensible distribution $\Lambda(\{q_{k}\})$. The condition for obtaining a {\em linear}
SSE is we have to be able to write the time derivative of equation
(\ref{LMeasurementSystemState}) in terms of only $\lpsiqk{t}$.

A linear SSE is only really useful if it can be transformed into a nonlinear SSE for the
normalized state $\psiqk{t}$. To do this one requires that there exists a Girsanov
transformation for the variables $\{q_{k}\}$ \cite{GatGis91}. This is a transformation
that takes into account the relation between the actual probability to ostensible
probability,
\begin{equation}\label{Girsanov}
{P}(\{q_{k}\},t)=\lpsiqb{t}\tilde{\psi}_{\{q_{k}\}}(t){\rangle}
\Lambda({\{q_{k}\}}),
\end{equation}
which follows from Eqs.~(\ref{LMeasurementSystemState}) and
(\ref{Probability}). Specifically, the Girsanov transformation is
a time-dependent transformation that changes the variables
$\{q_{k}\}$ into the variables $\{{q}_{k}^{\Lambda}\}$ such that
\beq \label{Girs1}
\Lambda(\{q_{k}^{\Lambda}\})d\{q_{k}^{\Lambda}\} =
P(\{q_{k}\},t)d\{q_{k}\}. \eeq

We can see the usefulness of this transformation as follows. If we
normalize the unnormalized states, but keep the same ostensible
distribution, then the ensemble average will not reproduce
$\rho_{\rm red}(t)$: \beq \int \frac{\Lambda
(\{q_k\})}{|\tilde{\psi}_{\{q_k\}}|^{2}}\lpsiqk{t}\lpsiqb{t}
d\{q_k\} \neq \rho_{\rm red}(t). \eeq However if $\{q_{k}\}$ are
chosen from the actual distribution then, of course it does: \beq
\int \frac{P
(\{q_k\},t)}{|\tilde{\psi}_{\{q_k\}}|^{2}}\lpsiqk{t}\lpsiqb{t}
d\{q_k\} = \rho_{\rm red}(t). \eeq Equivalently, using the
ostensible distribution for $\{q_{k}^{\Lambda}\}$, \beq \int
\frac{\Lambda
(\{q_k^{\Lambda}\})}{|\tilde{\psi}_{\{q_k\}}|^{2}}\lpsiqk{t}\lpsiqb{t}
d\{q^{\Lambda}_k\} = \rho_{\rm red}(t). \eeq Note that both
$\{q_{k}\}$ and $\{q_{k}^{\Lambda}\}$ appear here. This means that
if we have a linear SSE, we can derive a nonlinear (`actual') SSE
by normalizing the state
\begin{equation}\label{MeasurementSystemState2}
\psiqk{t}=\frac{1}{|\tilde{\psi}_{\{q_k\}}|}\lpsiqk{t},
\end{equation} where
\begin{equation}\label{norm}
{|\tilde{\psi}_{\{q_k\}}|}=\rt{\lpsiqb{t}\tilde{\psi}_{\{q_{k}\}}(t){\rangle}},
\end{equation}
but generating the SSE by drawing $\{q_{k}^{\Lambda}\}$ rather than $\{q_{k}\}$
from the ostensible distribution.

Now that we know how to use Eq.~(\ref{MeasurementSystemState2}), we can calculate the time
derivative of $\psiqk{t}$ in terms of $\lpsiqk{t}$. This results in
\begin{equation}\label{diffSystemState}
d_t\psiqk{t}=\frac{1}{|\tilde{\psi}_{\{q_k\}}|} d_t\lpsiqk{t}+\lpsiqk{t}d_t
\frac{1}{|\tilde{\psi}|},
\end{equation} where
\begin{equation} \label{diffTerm1}
d_t\lpsiqk{t}={\partial_t}\lpsiqk{t}+\sum_{k}d_t{q}_{k}{\partial_{q_{k}}}\lpsiqk
{t}.
\end{equation}
Here we have assumed that we can define $d_{t} q_{k}$ so as to generate a $q_{k}(t)$ which
ensures that \erf{Girs1} is always satisfied. From the above discussion, it is thus
apparent that three conditions must be satisfied if \erf{diffSystemState} is to be a SSE
for the system state $\psiqk{t}$. These are:

1. It is possible to obtain a linear SSE, that is
${\partial_t}\lpsiqk{t}$.

2. There is a Girsanov transformation $\{q_k^{\Lambda}\} \to \{q_k(t)\}$ such that an
equation for $d_t{q}_{k}$ for all $k$ can be found explicitly.

3. Equation (\ref{diffSystemState}) can be written in terms of only
$\psiqk{t}$.

If we can satisfy all these conditions then we have a SSE which
generates a state with a definite physical interpretation.
The SSE generates a state at time $t$ which is of the form of
Eq.~(\ref{MeasurementSystemState}). This is clearly the normalized
state conditioned on a measurement being
performed at time $t$ on the entire bath, and yielding results
$\{q_{k}\}$.

It is important to note, however, that the
linking of the states at earlier times to form a trajectory (which
is how the SSE generates the state at time $t$) appears to be
 a convenient fiction. A measurement on the whole bath at time $t$ is
 clearly incompatible with a similar measurement at an earlier time.
It is only in the Markovian limit that compatible bath
measurements can be made, so that the quantum trajectory as a
whole can be interpreted physically.  In other words the time
evolution generated by the SSE simply links together hypothetical
conditioned states at different times, with different measurement
results $\{q_{k}(t)\}$. The relation between the results at
different times is purely mathematical, not physical. The
mathematical relation comes from the time-dependent Girsanov
transformation: the $q^{\Lambda}_{k}$ corresponding to the
$q_{k}(t)$ are the same at all times.

\section{Coherent Bath Unraveling} \label{CoherentUnravelling}

\subsection{Coherent Noise Operator} \label{NMCoherentNoiseOperators}

The first unraveling we consider is that associated with the bath being
projected into a multitude mode coherent states, that is  $\qk=\ak$ where
\begin{equation}
  \ak=\prod_{k} \frac{1}{\rt{\pi}}e^{-|a_{k}|^{2}/2}\sum_{n_{k}}
   \frac{a_{k}^{n_{k}}}{\rt{n_{k}!}}\ket{n_{k}}.
\end{equation}
Note that these states are deliberately not normalized, so that the
multi-mode integral of the
effect $\hat F_{\{a_{k}\}}=\ak \bra{\{a_{k}\}}$ is unity.
We call the resultant unraveling the `coherent state unraveling'.
For this unraveling we define the noise
operator
\begin{equation} \label{CNoiseOperator}
\hat{z}(t)=\hat{b}(t)e^{i\omega_{0}
t}=\sum_{k}g_{k}\hat{a}_{k}e^{-i\Omega_{k} t},
\end{equation} where $\Omega_{k}=\omega_{k}-\omega_{0}$. This noise
operator has the property
\begin{equation}
\hat{z}(t)\ak=z(t)\ak,
\end{equation} where $z(t)$ is the noise
function,  given by
\begin{equation} \label{CNoiseFunction}
{z}(t)=\sum_{k}g_{k}{a}_{k}e^{-i\Omega_{k} t}.
\end{equation}

An important property of the bath is its correlation: how the noise operator (function) at
time $t$ is related to that at time $s$. This is determined  by the commutator (operators)
or correlation function (noise functions). For a non-Hermitian operator there are two
important commutators,
\begin{eqnarray} \label{CCommutator1}
    [\hat{z}(t),\hat{z}(s)]&=& 0,\\\label{CCommutator2}
[\hat{z}(t),\hat{z}(s)\dg]&=&\alpha{(t-s)},
\end{eqnarray}
where, in the notation of DSG,
\begin{equation}\label{MemoryFunction}
  \alpha{(t-s)}=\sum_{k}|g_{k}|^{2}e^{-i\Omega_{k} (t-s)},
\end{equation}
which we call the memory function.

The second form of correlation in defined in terms of the noise functions as ${\rm
E}[z(t)z^{*}(s)]$. This depends on the probability for obtaining the results $\{a_{k}\}$
in the measurement at the two times. In linear QMT, these probabilities are given by the
ostensible distribution $\Lambda(\{a_{k}\})$, which may be chosen to be time-independent.
It is convenient to choose $\Lambda(\{a_{k}\})$ to be equal to the actual probability that
would arise when the bath is always in the vacuum state. That is,
\begin{equation} \label{CProbability}
{\Lambda}(\{a_{k}\})=\langle{\{0_k\}}\ak\ab{\{0_k\}}\rangle
={\pi^{-\kappa}}{e^{-\sum_{k}|a_{k}|^{2}}},
\end{equation} where $\kappa=\sum_{k}$.
As will be seen later, this is appropriate if the bath is initially in
this state. The correlation for the noise functions under this assumption is,
\begin{eqnarray} \label{CCorrelation1}
 \tilde{ {\rm E}}[{z}(t)z^{*}(s)]&=&\alpha{(t-s)},\\ \label{CCorrelation2}
 \tilde{ {\rm E}}[{z}(t)z(s)]&=&0.
\end{eqnarray}
Note that we have used the notation discussed below \erf{deftildeE}.
Thus for the special case where the ostensible probability is given
by Eq.~(\ref{CProbability}),
the memory function is equal to
the correlation of the noise functions.

\subsubsection{The Markov Limit}

Since one of our aims is to consider the Markovian limit of our non-Markovian SSEs (in
which one obtains a genuine quantum trajectory), the Markov limit of all our main results
will be presented. In the Markov limit the number of modes become continuous and the
coupling constant $|g_k|$ becomes flat ($|g_k|=g$) and equal to $\rt{\gamma/2\pi}$. This
allows us to write
\begin{eqnarray}
\alpha{(t-s)}&=&\frac{\gamma}{2\pi}\int_{0}^{\infty}
e^{-i(\omega-\omega_0)(t-s)}d\omega
\nn \\ &=&\frac{\gamma}{2\pi}\int_{-\omega_0}^{\infty}
e^{-i\Omega(t-s)}d\Omega,
\end{eqnarray} and for optical situations (high $\omega_{0}$ situations)
 with little error this can be written as
\begin{equation}
\alpha{(t-s)}=\frac{\gamma}{2\pi}\int_{-\infty}^{\infty}
e^{-i\Omega(t-s)}d\Omega=\gamma\delta{(t-s)}.
\end{equation} Therefore,
\begin{eqnarray} \label{MCCommutator1}\label{MCCorrelation1}
 \tilde{\rm E}[z(t)z^{*}(s)] \,=\,
 [\hat{z}(t),\hat{z}(s)\dg]&=&\gamma\delta{(t-s)}, \\\label{MCCommutator2}
\tilde{{\rm E}}[{z}(t)z(s)]\,=\, [\hat{z}(t),\hat{z}(s)]&=&0.
\end{eqnarray}
 This implies that ostensibly
 $z(t)$ is a complex gaussian random variable (GRV) of mean 0
and variance $\gamma/ dt$. That is, $z(t)=\rt{\gamma}\zeta(t)$, where $\zeta(t)$ is the
standard complex white noise function \cite{Gar83}. These are the correct correlation
function for the heterodyne noise functions \cite{Wis96}.

\subsection{The Linear Stochastic \sch Equations for the Coherent
Unraveling} \label{CoherentLSSE}

In this section we will derive the linear non-Markovian SSEs for the
ostensible
probability introduced above, and show that in the Markov limits it
gives the linear Heterodyne SSE.  We use many of the same techniques
as DSG.
To calculate the linear SSE we write the \sch equation in terms of the
noise operator,
$\hat{z}(t)$
\begin{equation}
d_{t}\ket{\Psi(t)}=\Big{\{}
{-i}\hat{H}(t)+\hat{z}\dg(t)\hat{L}-\hat{z}(t)\hat{L}\dg\Big{\}}\ket{\Psi(t)}.
\end{equation}
Then by differentiating equation (\ref{LMeasurementSystemState})
with respect to time (with $q_{k}$ set to $a_{k}$) we obtain
\begin{eqnarray} \label{step}
  \partial_{t}\lpsiak{t}&=&\Big{\{}
{-i}\hat{H}(t)+z^{*}(t)\hat{L}\Big{\}}\lpsiak{t}\nl-\frac{\ab\hat{z}(t)\hat{L}
\dg\ket{\Psi(t)}} {\rt{\Lambda({\{a_{k}\}})}},
\end{eqnarray} as $\hat H(t)$ is a system-only operator and $\ab$
is the left-eigenstate of $\hat{z}(t)\dg$.
To satisfy the condition for a linear SSE we must evaluate the
last term in this equation in terms of $\lpsiak{t}$. To do this we
use \cite{ScuZub97},
\begin{equation}
 \ab \hat{a}_{k}\ket{\Psi(t)}= \Big{(} \frac{a_{k}}{2}+{\partial_{
  a^{*}_{k}}}\Big{)}\ab{\Psi(t)}\rangle
\end{equation} and
\begin{equation}
{\partial_{a^{*}_{k}}}\lpsiak{t}=\frac{\partial_{a^{*}_{k}}
\ab{\Psi(t)}\rangle}{\rt{
\Lambda({\{a_{k}\}})}}+\frac{a_{k}}{2}\lpsiak{t}.
\end{equation} With these two expressions and the definition of $\hat{z}(t)$,
\begin{equation}
\frac{\ab \hat{z}(t)\ket{\Psi(t)}}{\rt{ \Lambda({\{a_{k}\}})}}= \sum_{k}g_k
e^{-i
\Omega_{k}t} {\partial_{a^{*}_{k}}}\lpsiak{t}.
\end{equation}
This allows us to write equation (\ref{step}) as
\begin{eqnarray} \label{CLSSEPartialForm}
  \partial_{t}\lpsiak{t}&=&\Big{\{}
{-i}\hat{H}(t)+z^{*}(t)\hat{L}- \hat{L}\dg\sum_{k}g_k e^{-i
\Omega_{k}t}\nl\times {\partial_{a^{*}_{k}}}\Big{\}}\lpsiak{t}.
\end{eqnarray}

This is a linear equation in terms of $\{a_{k}\}$. Note that it is
not really a SSE, as the final term implies that the evolution of
the state $\lpsiak{t}$ depends not only on itself, but upon
neighbouring states with different values of $\{a_{k}\}$. That is,
we cannot simply choose (stochastically) a value for $\{a_{k}\}$
from the ostensible distribution and then propagate forward the
system state using that value. However, we can make progress
towards an equation where we can do this by rewriting the partial
derivative in terms of a functional derivative. This is done by
using the following relation (see for example \cite{ParYan89}),
\begin{equation} \label{PartialFunctional}
{\partial_{a^{*}_{k}}}=\int_{0}^{t}\frac{\delta}{\delta
z^{*}(s)}\frac{\partial
z^{*}(s)}{\partial {a^{*}_{k}}} ds,
\end{equation} where $0$ is the initial time. This gives
\begin{eqnarray} \label{CLSSEFunctionalForm}
  \partial_{t}\lpsizk{t}&=&\Big{\{}
{-i}\hat{H}(t)+z^{*}(t)\hat{L}-\hat{L}\dg
\int_{0}^{t}\alpha{(t-s)}\nl\times\frac{\delta}{\delta z^{*}(s)}
ds\Big{\}}\lpsizk{t},
\end{eqnarray} where $\alpha{(t-s)}$ is defined in equation
(\ref{MemoryFunction}). By replacing the partial derivatives by the functional derivative
we have enforced the initial condition $\ket{\Psi(0)}=\ket{\{0_{k}\}}\ket{\psi(0)}$,
 This is seen as follows.
 At $t=0$ the functional derivative term in the above equation will have zero
contribution, from the definition (\ref{PartialFunctional}). By
comparison with the corresponding term in \erf{CLSSEPartialForm}, it
follows that  $\partial_{a^{*}_{k}}\lpsiak{t}|_{t=0}=0$ for all $k$. From
\erf{LMeasurementSystemState} this is only possible if
the system and bath states initially (at time $0$) factorize, and
if $\Lambda(\{a_{k}\})=|\langle\{a_{k}\}|\psi_{\rm bath}\rangle|^2$.
From our choice (\ref{CProbability}) of ostensible probability, this enforces
$\ket{\psi_{\rm bath}}=\ket{\{0_{k}\}}$. This is physically acceptable as we may assume
that at time $0$ the system and bath are uncoupled, and the bath is in the vacuum state.

Like \erf{CLSSEPartialForm}, \erf{CLSSEFunctionalForm} is not really a SSE because the
functional derivative means that it depends not upon a state $\lpsizk{t}$ at all times for
a single value of the function $z(t)$, but rather also upon states for other values of
that function. That is, we cannot stochastically choose $z(t)$ in order to generate a
trajectory independent of other trajectories. Instead, all possible trajectories would
have to be calculated in parallel. This means that the amount of calculation involved in
solving \erf{CLSSEFunctionalForm} would be comparable to that required for directly
solving the \sch equation (\ref{SchEquation}). However in some circumstance we can make
the following Ansatz \cite{DioGisStr98},
\begin{equation}\label{Ansatz}
  \frac{\delta}{\delta
  z_{\phi}(s)}\lpsizk{t}=\hat{O}_{z}(t,s)\lpsizk{t},
\end{equation}
where $\hat{O}_{z}(t,s)$ is some system operator which is a function
of $t$, and $s$, and a functional of $z$.
With this Ansatz the linear SSE becomes
\begin{eqnarray} \label{CLSSEOperatorForm}
  \partial_{t}\lpsizk{t}&=&\Big{\{}
{-i}\hat{H}(t)+z^{*}(t)\hat{L}-\hat{L}\dg\int_{0}^{t}
\alpha{(t-s)} \nl\times\hat{O}_{z}(t,s) ds\Big{\}}\lpsizk{t}.
\end{eqnarray}
This is now a true SSE, where each trajectory can be evolved
independently. It is the same as the linear SSE that DSG presented
in Ref.~\cite{DioStr97,DioGisStr98}. Note that it is non-Markovian
because the noise $z^{*}(t)$ is non-white, because of the finite
lower limit of the integral, and because $\hat{O}_{z}(t,s)$ may
depend upon $z$.

\subsubsection{The Markov Limit}

The next question is what is the Markov limit of this equation? To find this we use the
results of section (\ref{NMCoherentNoiseOperators}) and the fact that
$\hat{O}_{z}(t,t)=\hat{L}$ \cite{DioGisStr98}. Applying them to equation
(\ref{CLSSEOperatorForm}) results in
\begin{equation} \label{MCLSSEStratonovichForm}
 \partial_{t}\lpsizk{t}=\Big{\{}
{-i}\hat{H}(t)+z^{*}(t)\hat{L}-\gamma \hat{L}\dg\hat{L}\Big{\}}\lpsizk{t},
\end{equation} where $z(t)=\rt{\gamma}\zeta(t)$.
By its method of derivation, this equation is in Stratonovich form \cite{Gar83}. To
compare with the standard Markov equations we should convert it to an \ito SSE. This can
be derived by using an arbitrary basis and defining $\psi_j=\langle j \ket{\psi}$ and
$L_{j,k}=\bra{j}\hat{L}\ket{k}$. Then if the \str form is \beq
\partial_{t}{\psi}_{j} = a_j+b_j \zeta^{*}(t) ,
\eeq
the \ito form (which we indicate by use of the infinitesimals rather
than the derivatives)  is
\beq
\label{ItoStrZeta} \label{CorrectionTermCVCN}
d\psi_j(t) = a_j dt+ b_j d\zeta^{*}(t)dt +\frac{d t}{2}\sum_{l} b_l^{*}
\frac{\partial}{\partial\psi_l^*}b_j.
\eeq
The final term here is the \ito correction term.
Looking at equation (\ref{MCLSSEStratonovichForm}) we see
that, $b_{j}=\rt{\gamma}\sum_{k}L_{j,k}\psi_k$, and since
${\partial \psi_k}/{\partial\psi_l^*}$ is zero for all $k$,
the correction term for this equation is 0. Thus the \ito SSE is,
\begin{equation} \label{MLSSEitoForm}
d\lpsizk{t}=dt\bl-i\hat{H}(t)+\hat{L}z^{*}(t)
  -\frac{\gamma}{2}\hat{L}\dg\hat{L}\br\lpsizk{t},
\end{equation} which is the standard linear heterodyne SSE presented in
Ref.~\cite{GoeGra94} as $z(t)=\rt{\gamma}\zeta(t)=\rt{\gamma}(\xi_1(t)+i\xi_2(t))$,
where $\xi_{k}(t)$ are the standard real-valued white noise terms \cite{Gar83}.
\label{sec:itostr1}

\subsection{The Actual Stochastic \sch Equations for the Coherent Unraveling}

In this section we will derive the non-Markovian SSEs for the actual
probability
distribution and show
that in the Markov limits it gives the the usual heterodyne SSE.
Again, we use many of the same techniques
as DSG.

As discussed in section \ref{QMT}, to find an actual (i.e. nonlinear) SSE
for the normalized state we
need to satisfy 3 conditions. The first was to derive a linear SSE,
which we did in the preceding section (by making use of an anstaz).
The second
condition is to find random variables with the
actual probabilities of measurement results. To work out these random
variables,
$\{{a}_{k}\}$ we use the Girsanov transform (\ref{Girsanov})
to find a first-order partial differential equation (PDE) for the
probability, from which the characteristic equation generates the
transformed variables.

To obtain the PDE we differentiate \erf{Girsanov}, giving
\begin{equation}
\partial_{t}{ P} (\{a_k\},t)= \lpsiab{t} \partial_{t}
\lpsiak{t}\Lambda(\{a_k\})+{\rm c.c}.
\end{equation}
By equation (\ref{CLSSEPartialForm}) the
above becomes
\begin{eqnarray}
\partial_{t}{ P} (\{a_k\},t)&=&\Big{\{} \lpsiab{t} \hat{L}
\lpsiak{t}\sum_{k}a^{*}_{k}g_{k}^{*}e^{i\Omega_{k} t}\nl-
\sum_{k}\lpsiab{t} \hat{L}\dg
\partial_{a_{k}^{*}}\lpsiak{t}g_{k}e^{-i\Omega_{k}t}\nl+{\rm c.c.}
\Big{\}}\Lambda(\{a_k\}),
\end{eqnarray}
Using the fact that $\lpsiak{t}$ is analytical in $a_{k}^{*}$ (so that
$\partial_{a_{k}}\lpsiak{t}=0$) \cite{DioStr97}, and the product rule for differentiation,
we can simplify the above to
\begin{eqnarray}
\partial_{t}{ P}
(\{a_k\},t)&=&-\sum_{k}g_{k}e^{-i\Omega_{k}t}
 \partial_{a_{k}^{*}}\Big{\{}\lpsiab{t}
\hat{L}\dg\nl\times\lpsiak{t}\Lambda(\{a_k\})\Big{\}}+{\rm c.c}.
\end{eqnarray}
Defining
\begin{equation} \label{Expect}
\langle \hat{L}\dg\rangle_t=\psiab{t}
\hat{L}\dg\psiak{t}=\frac{\lpsiab{t}
\hat{L}\dg\lpsiak{t}}{\langle{\tilde{\psi}_{\{a_{k}\}}(t)}\lpsiak{t}}
\end{equation} allows us to write
\begin{eqnarray} \label{CoherentFokkerPlank}
\partial_{t}{ P}
(\{a_k\},t)&=&-\sum_{k}g_{k}e^{-i\Omega_{k}t}
 \partial_{a_{k}^{*}}\Big{\{}\langle
\hat{L}\dg\rangle_t{P} (\{a_k\},t)\Big{\}}\nl+{\rm c.c.}
\end{eqnarray} This is the PDE for the
probability distribution.

At $t=0$, we have from \erf{Girsanov} that
\begin{equation}
{ P} (\{a_k\},0)=
\langle{\tilde{\psi}_{\{a_{k}\}}(0)}\lpsiak{0}\Lambda(\{a_k\}).
\end{equation}
As noted above, to obtain \erf{CLSSEFunctionalForm}  we had to assume that the bath was
initially in the vacuum state, uncorrelated with the system. This enforces the equation of
the initial probability distribution to be the ostensible distribution
\begin{equation} \label{CProbPart}
{ P} (\{a_k\},0)=\Lambda(\{a_k\})=
{\pi^{-\kappa}}{e^{-\sum_{k}|a_{k}|^{2}}}.
\end{equation}

From this PDE we can find the characteristic equations
\begin{equation}\label{CoherentRandomDiff}
  d_{t}a_{k}^{*}=g_{k}e^{-i\Omega_{k}t}\langle \hat{L}\dg\rangle_{t},
\end{equation} which integrates to give
\begin{equation}\label{CoherentRandomDVariable}
  a_{k}^{*}(t)=a_{k}^{*}(0)+\int_{0}^{t} g_{k}e^{-i\Omega_{k}s}\langle
\hat{L}\dg\rangle_s
  ds.
\end{equation}
The random variable $a_{k}^{*}(0)$ is one with probability
distribution  (\ref{CProbPart}). With equation
(\ref{CoherentRandomDVariable}) and our noise function definition,
Eq.~(\ref{CNoiseFunction}), we can write $z(t)$ as
\begin{equation}
  z^{*}(t)=a_{k}^{*}(0)g_{k}^{*} e^{i\Omega_{k}t}+\int_{0}^{t}
\alpha^{*}{(t-s)} \langle \hat{L}\dg\rangle_s
  ds.
\end{equation}
The term $a_{k}^{*}(0)g_{k}^{*}e^{i\Omega_{k}t}$ is the noise
function one would obtain if the bath were assumed to be in the
vacuum state. This is our assumption for the ostensible
distribution so we will label this term $z^{*}_{\Lambda}(t)$. This
allows us to write
\begin{equation} \label{CRealNoiseOperator}
  z^{*}(t)=z_{\Lambda}^{*}(t)+\int_{0}^{t} \alpha^{*}{(t-s)} \langle
\hat{L}\dg\rangle_s  ds,
\end{equation} where $z^{*}_{\Lambda}(t)$ obeys the correlations expressed in
equations (\ref{CCorrelation1}) and (\ref{CCorrelation2}).

The third condition was to show that we can write equation (\ref{diffSystemState}) in
terms of only $\psizk{t}$. To do this we start by calculating $d_t\lpsizk{t}$. Using
equations (\ref{diffTerm1}), (\ref{CLSSEFunctionalForm}) and (\ref{PartialFunctional}) we
get,
\begin{eqnarray}
d_t\lpsizk{t}&=&\Big{\{}
{-i\hat{H}(t)}+z^{*}(t)\hat{L}-(\hat{L}\dg-\langle\hat{L}\dg\rangle_t)
\nl\times\int_{0}^{t}\alpha{(t-s)} \frac{\delta }{\delta z^{*}(s)
} ds \Big{\}}\lpsizk{t}.
\end{eqnarray}
Looking at Eq.~(\ref{diffSystemState}) we see that to obtain the
actual SSE we need to calculate $\lpsizk{t}d_t
{|\tilde{\psi}_{\{a_k\}}|}^{-1}$. Using the above,
\begin{widetext}
\begin{eqnarray}
\lpsizk{t}d_t
\frac{1}{|\tilde{\psi}_{\{a_k\}}|}&=&-\frac{\psizk{t}}{|\tilde{\psi}_{\{a_k\}}|}
(\psizb{t}d_t\lpsizk{t}+{\rm c.c}) \nn \\&=&-\Big{\{}
z^{*}(t)\langle\hat{L}\rangle_t-\psizb{t}(\hat{L}\dg-\langle\hat{L}\dg\rangle_t)
\frac{1}{|\tilde{\psi}_{\{a_k\}}|}\int_{0}^{t}\alpha{(t-s)}
\frac{\delta }{\delta z^{*}(s) }\lpsizk{t}ds\nl+{\rm c.c}
\Big{\}}\psizk{t}/2.
\end{eqnarray}
Therefore Eq.~(\ref{diffSystemState}) becomes
\begin{eqnarray}
d_t\psizk{t}&=&\Big{\{}
{-i\hat{H}(t)}+z^{*}(t)\hat{L}\Big{\}}\psizk{t}-(\hat{L}\dg-\langle\hat{L}\dg\rangle_t)
\frac{1}{|\tilde{\psi}_{\{a_k\}}|}\int_{0}^{t}\alpha{(t-s)}
\frac{\delta }{\delta z^{*}(s) }\lpsizk{t}ds\nl -\psizk{t}\Big{\{}
z^{*}(t)\langle\hat{L}\rangle_t-\psizb{t}(\hat{L}\dg-\langle\hat{L}\dg\rangle_t)
\frac{1}{|\tilde{\psi}_{\{a_k\}}|}\int_{0}^{t}\alpha{(t-s)}
\frac{\delta }{\delta z^{*}(s) }\lpsizk{t}ds\nl+{\rm
c.c}\Big{\}}/2.
\end{eqnarray}\end{widetext}

This can be simplified by using the fact that if our SSE has the
form $d_t\ket{\psi}=(\hat{A}+B/2+B^{*}/2)\ket{\psi}$ then we can
define a state $\ket{\phi}=\exp(\int(B-B^{*})dt/2)\ket{\psi}$
(which is the same state as $\ket{\psi}$) that gives a equivalent
SSE, of form $d_t\ket{\phi}=(\hat{A}+B)\ket{\phi}$. Applying this
to the above gives \begin{widetext}
\begin{eqnarray} \label{CSSEFunctionalForm}
d_t\psizk{t}&=&\Big{\{}
{-i\hat{H}(t)}+z^{*}(t)(\hat{L}-\langle\hat{L}\rangle_t)\Big{\}}\psizk{t}-(\hat{
L}\dg-\langle\hat{L}\dg\rangle_t)
\frac{1}{|\tilde{\psi}_{\{a_k\}}|}\int_{0}^{t} \alpha{(t-s)}
\frac{\delta}{\delta
z^{*}(s)} ds \lpsizk{t}\nl+\psizk{t}
\psizb{t}(\hat{L}\dg-\langle\hat{L}\dg\rangle_t)
\frac{1}{|\tilde{\psi}_{\{a_k\}}|}\int_{0}^{t} \alpha{(t-s)}
\frac{\delta}{\delta
z^{*}(s)} ds \lpsizk{t}.
\end{eqnarray}\end{widetext}
This is not yet a SSE as it still contains $\lpsizk{t}$ terms, however if we can make the
Ansatz described by \erf{Ansatz} we can write this as
\begin{eqnarray}\label{CSSEOperatorForm}
d_t\psizk{t}&=&\Big{[}
{-i\hat{H}(t)}+z^{*}(t)(\hat{L}-\langle\hat{L}\rangle_t)\nl-\int_{0}^{t}
\alpha{(t-s)}\Big{\{}(\hat{L}\dg-\langle\hat{L}\dg\rangle_t)
 \hat{O}_{z}(t,s) \nl-
\Big{\langle}(\hat{L}\dg-\langle\hat{L}\dg\rangle_t)
 \hat{O}_{z}(t,s)  \Big{\rangle}_t\Big{\}}ds
\Big{]}\psizk{t},\nl
\end{eqnarray}
which is a genuine SSE. This means that an actual SSE (generating
normalized states with their actual probabilities) can only be
found if we can make the
 Ansatz describe in Eq.~(\ref{Ansatz}).

This SSE is the same
as that presented in Refs.~\cite{DioGisStr98,StrDioGis99}. As shown
here, it gives us the state the system
would be in if at time $t$ we performed a measurement in the
coherent basis, and the result was $z(t)$ as defined in
\erf{CRealNoiseOperator}.
Note that this means that the result $z(t)$ depends upon the system
state at earlier times in the trajectory generated by the above SSE.
We have argued above that this linking of states at different times
is a convenient fiction, but we see here that it is mathematically
necessary in order to generate measurement results for a particular
time with the actual probability.

\subsubsection{The Markov Limit}

Finally, we are again interested in the Markov limit of this SSE. Taking
the Markov limit of the noise function, one obtains
\begin{equation} \label{MCNoiseFunction}
  z^{*}(t)=z_{\Lambda}^{*}(t)+\frac{\gamma}{2} \langle
  \hat{L}\dg\rangle_t,
 \end{equation} where $z_{\Lambda}^{*}(t)=\rt{\gamma}\zeta^{*}(t)$.

To apply the Markov limit to \erf{CSSEOperatorForm} we use
$\alpha{(t-s)}\rightarrow\gamma\delta{(t-s)}$ and
$\hat{O}_z(t,t)=\hat{L}$, resulting in
\begin{eqnarray}\label{MCSSEStratonovichForm}
d_t\psizk{t}&=&\Big{\{}
\frac{-i\hat{H}(t)}{\hbar}+(\hat{L}-\langle\hat{L}\rangle_t)(z^{*}(t)+\frac{\gamma}{2}\langle
\hat{L}\dg\rangle_t)\nl-\frac{\gamma}{2}(\hat{L}\dg\hat{L}-\langle\hat{L}\dg\hat
{L}\rangle_t) \Big{\}}\psizk{t},
\end{eqnarray} which is in Stratonovich form. To convert this to an \ito
SSE we have
to calculate the
\ito correction term in Eq.~(\ref{CorrectionTermCVCN}). For
this equation, the correction term is
\begin{equation}
  \frac{dt \gamma}{2}\Big{(}-\langle\hat{L}\dg\hat{L}\rangle_t
  +\langle\hat{L}\dg\rangle_t\langle\hat{L}\rangle_t\Big{)}\psizk{t},
\end{equation} which with Eq. (\ref{MCSSEStratonovichForm}) results in,
\begin{eqnarray}\label{MCSSEItoForm}
d_t\psizk{t}&=&\Big{\{}
\frac{-i\hat{H}(t)}{\hbar}+(\hat{L}-\langle\hat{L}\rangle_t)z^{*}(t)\nl-
\frac{\gamma}{2}(\hat{L}\dg\hat{L}-\hat{L}\langle
\hat{L}\dg\rangle_t) \Big{\}}\psizk{t}.
\end{eqnarray}
This is the \ito SSE for the actual measurement probabilities. When we substitute in
$z^{*}(t)$ from \erf{MCNoiseFunction} we get the same heterodyne SSE as that presented in
Ref.~\cite{GisPer93,WisMil93c}.

Readers familiar with quantum trajectory theory for heterodyne
detection may be puzzled by the factor of $1/2$ multiplying the
deterministic contribution to $z(t)$. This function is, according
to the above theory, the result of measuring the bath at time $t$ in
the coherent state basis.
But in the usual quantum trajectory theory \cite{WisMil93c} the
measured (complex) heterodyne current at time $t$ is
\beq
I(t) = \rt{\gamma}\zeta(t)+{\gamma}\langle\hat{L}\rangle_t.
\eeq
which lacks the $1/2$. Where does this discrepancy come from?
 To answer this question we have to consider
the definition of a measurement, and in particular the time of the measurement. In quantum
trajectory theory we must consider the measurement which conditions the state at time $t$
as actually occurring at a time $t+dt$ \cite{Wis96}. That is, the $\delta$-correlated bath
must be given a chance to interact with the system before the measurement is made. By
contrast, in the above theory the measurement occurs exactly at time $t$. For a
non-Markovian bath (with a finite correlation time) the difference between $t$ and $t+dt$
is infinitesimal. However in the Markov limit, this infinitesimal difference in
measurement time causes the finite difference between $z(t)$ and $I(t)$.

It is easiest to see this using the Heisenberg picture.
From the above theory,
\begin{eqnarray}
{\rm E}[z(t)]&=&\bra{\Psi(t)}\hat{z}(t)\ket{\Psi(t)}\nn \\ &=&
\bra{\psi(0)}\bra{\{0_{k}\}}U_{\rm int}\dg(t)\hat{z}(t)U_{\rm int}(t)\ket{\{0_{k}\}}\ket{\psi(0)} \nn\\
& =&\bra{\psi(0)}\bra{\{0_{k}\}}\hat{z}_{H}(t)\ket{\{0_{k}\}}
\ket{\psi(0)},
\end{eqnarray} where $\hat{z}_{H}(t)$ is the Heisenberg noise
operator. In quantum trajectory theory the measurement is defined to take
place after the system and bath have interacted for a time $dt$, so
that
\begin{eqnarray}
{\rm E}[I(t)]&=&\bra{\Psi(t+dt)}\hat{z}(t)\ket{\Psi(t+dt)}\nn \\
&=&
\bra{\psi(0)}\bra{\{0_{k}\}}U_{\rm int}\dg(t+dt)\hat{z}(t)U_{\rm int}(t+dt)\nl\times\ket{\{0_{k}\}}\ket{\psi(0)}\nn\\
& =&\bra{\psi(0)}\bra{\{0_{k}\}}\hat{I}(t)\ket{\{0_{k}\}}
\ket{\psi(0)}.
\end{eqnarray} Therefore,
\begin{eqnarray}
  \hat{I}(t)&=&U_{\rm int}\dg(t+dt,t)\hat{z}_{H}(t)U_{\rm int}(t+dt,t).
\end{eqnarray}

By using standard Heisenberg equations it can be shown that
\begin{equation}
  \hat{I}(t)=\hat{z}_{H}(t)+\int_{t}^{t+dt}\alpha(t-s) U_{\rm int}\dg(s)\hat{L}
  U_{\rm int}(s)ds,
\end{equation} which has a Markov limit of the form
\begin{equation}
  \hat{I}(t)=\hat{z}_{H}(t)+\frac{\gamma}{2} U_{\rm int}\dg(t)\hat{L}
  U_{\rm int}(t).
\end{equation}
This is the operator form of the Heterodyne current, and shows the
extra contribution discussed above. It is similarly easy to show that
the  Markov form of
$\hat{z}_{H}(t)$ is
\beq
\hat{z}_{\Lambda}(t)+\frac{\gamma}{2}
 U_{\rm int}\dg(t)\hat{L}
  U_{\rm int}(t),
\eeq
where $\hat{z}_{\Lambda}(t)=\sum_{k}g_{k}\hat{a}_{k}(0)e^{-i\Omega_{k}t}$.
These relations are analogous to the Markovian input-output theory of Gardiner and
Collett \cite{GarCol85}. The correspondences are as follows
\bqa
{\hat z}_{\Lambda}(t) &\leftrightarrow& {\hat b}_{\rm in}(t), \\
{\hat z}_{H}(t) &\leftrightarrow& {\hat b}(t) ,\\
{\hat I}(t) &\leftrightarrow& {\hat b}_{\rm out}(t) .
\eqa

\section{Quadrature Bath Unraveling} \label{QuadratureUnravelling}

In this section we will present a second unraveling which is conditioned
on real noise
and has homodyne detection as its Markov limit.

\subsection{Quadrature Noise Operator} \label{quadOperatorSection}

To obtain a SSE with real noise, it is natural to consider a
 quadrature noise
operator,\begin{equation} \label{QuadratureNoiseOpeator}
\hat{z}(t)={\hat{b}(t)e^{i\omega_{0}
t}e^{-i\phi}+\hat{b}\dg(t)e^{-i\omega_{0} t}e^{i\phi}}
,\end{equation}
where $\hat{b}(t)$ is defined in equation
(\ref{BathOperatorInt}) and $\phi$ is some arbitrary phase.
The noise operator has a two-time commutator
\begin{eqnarray} \label{QCommuatorFull}
[\hat{z}(t),\hat{z}(s)]=\alpha{(t-s)}-\alpha^{*}{(t-s)},
\end{eqnarray}
independent of $\phi$. The
phase $\phi$ defines the measured quadrature: an
$x$-quadrature measurement occurs when  $\phi$ is set to zero, and the
conjugate measurement of the $y$-quadrature occurs when $\phi=\pi/2$.
Unless otherwise stated we will set $\phi$ to zero.

The basis for
the bath measurement is $\qk$ and must satisfy
\begin{equation}
  \hat{z}(t)\qk=z(t)\qk.
\end{equation}
The problem  with this noise function is that it is hard (maybe
impossible) to work out a time-independent eigenstate $\qk$ in the
interaction picture. However, we can find the eigenstate if we
make the assumptions that for every mode $k$ there exists another
mode, which we can label $-k$, such that $\Omega_{-k}=-\Omega_{k}$
and $g_{-k}=g^{*}_{k}$. These assumptions simply mean that the
modes coupled to the system come in symmetric pairs about the
system frequency $\omega_{0}$. Without loss of generality we can
take the $g_{k}$'s to be real, absorbing any phases in the
definitions of the bath operators. With all of these assumptions
we can rewrite equation (\ref{QuadratureNoiseOpeator}) as
\begin{equation}\label{QuadratureNoiseOperatorXY}
  \hat{z}(t)=\sum_{k>0} {2} g_{k}\bl  \hat{X}_{k}^{+}
  \cos(\Omega_{k}t)+\hat{Y}_{k}^{-}\sin(\Omega_{k}t)\br.
\end{equation}
Here we have introduced the two-mode quadrature operators
\begin{eqnarray}
\hat{X}_{k}^{\pm}&=&(\hat{x}_{k}\pm\hat{x}_{-k})/{\rt{2}}\label{X+}, \\
\hat{Y}_{k}^{\pm}&=&(\hat{y}_{k}\pm\hat{y}_{-k})/{\rt{2}}, \label{Y-}
\end{eqnarray}
where $\hat{x}_{k}$ and $\hat{y}_{k}$ are the quadratures of
$\hat{a}_{k}$:
\begin{equation} \label{aIintoxAndy}
  \hat{a}_{k}=({\hat{x}_{k}+i \hat{y}_{k}})/{\rt{2}}.
\end{equation}
These operators have the commutators
 \begin{eqnarray} \label{com1}
&&[\hat{X}_{k}^{-},\hat{Y}_{k}^{-}]=i, ~~~~~~
[\hat{X}_{k}^{-},\hat{Y}_{k}^{+}]=0,\\ \label{com2} &&
[\hat{X}_{k}^{+},\hat{Y}_{k}^{-}]=0,
~~~~~~ [\hat{X}_{k}^{+},\hat{Y}_{k}^{+}]=i.
\end{eqnarray}

Since $\{\hat{X}_{k}^{+}\}$ and $\{\hat{Y}_{k}^{-}\}$ form two
mutually commuting sets of commuting operators, and thus have a
common set of eigenstates. Since $\hat{z}(t)$ is a linear combination
of these operators, the eigenstates of
$\{\hat{X}_{k}^{+}\}$ and $\{\hat{Y}_{k}^{-}\}$
are  the $\qk$ we seek.
Therefore we can write the two eigenvalue equations,
\begin{eqnarray}
\hat{Y}_{k}^{-}\qk &=&{Y}_{k}^{-}\qk ,\label{EigenValueEquation2}\\
 \hat{X}_{k}^{+}\qk&=&{X}_{k}^{+}\qk. \label{EigenValueEquation1}
\end{eqnarray}
This suggest that we should write $\qk$ as $\ket{\{X_{k}^{+},Y_{K}^{-}\}}$, but for
brevity we will continue to write it as $\qk$. The form of the state that satisfies these
equations, in the $y_{k}$-basis, for a particular $k$ is
\begin{equation}\label{QState} 
\int
  \frac{dy'}{\rt{2\pi}}\ket{\lfrac{y'-Y_{k}^{-}}{\rt{2}}}_{-k}\ket{\lfrac
  {y'+Y_{k}^{-}}{\rt{2}}}_{k}e^{-iX_{k}^{+}y'},
\end{equation}
while in the $x_{k}$-basis it is
\begin{equation} \label{Qstatex}
\int \frac{dx'}{\rt{2\pi}}
  \ket{\lfrac{X_{k}^{+}-x'}{\rt{2}}}_{-k}\ket{\lfrac{X_{k}^{+}+x'}{\rt{2}
  }}_{k}e^{iY_{k}^{-}x'}.
\end{equation}

Under these assumption we can show that the memory function
$\alpha(t-s)$ in Eq.~(\ref{MemoryFunction})
becomes equal to the real function $\beta(t-s)$ given by
\begin{equation}\label{MemoryFunctionAssump}
\beta{(t-s)}=2\sum_{k>0}|g_{k}|^{2}\cos(\Omega_k (t-s)).
\end{equation} Thus the commutator expressed in Eq.~(\ref{QCommuatorFull})
becomes,
\begin{equation} \label{QCommuatorAssump}
[\hat{z}(t),\hat{z}(s)]=\beta{(t-s)}-\beta{(t-s)}=0.
\end{equation}
Moreover, the noise function is
\begin{equation} \label{QuadratureNoiseFunction}
z(t)=\sum_{k>0} {2} g_{k} \bl  X_{k}^{+}
  \cos(\Omega_{k}t)+Y_{k}^{-}\sin(\Omega_{k}t)\br.
\end{equation}
Since $X_{k}^{+}$ and $Y_{k}^{-}$ are
real, $z(t)$ is also.

We can define the correlation function for the noise functions as ${\rm E}[z(t)z(s)]$, and
again this depends on the probability distribution for the variables $X_{k}^{+}$ and
$Y_{k}^{-}$. It is again convenient to choose the ostensible distribution to be that
corresponding to the bath being in the vacuum state. Explicitly we then have \beq
\label{QuadratureProbability} \Lambda(\{X_{k},Y_{k}\}) = \pi^{-\kappa/2}{e^{-\sum_{k>0}
({X_{k}^{+}}^2+{Y_{k}^{-}}^2)}}. \eeq
With the usual ostensible distribution the correlation
function is
\begin{equation} \label{NMNoiseCorrelation}
\tilde{\rm E}[z(t)z(s)]=2\sum_{k>0}|g_{k}|^{2} \cos(\Omega_{k}(t-s))=\beta{(t-s)},
  \end{equation}
while $\tilde{\rm E}[z(t)]=0$ as before.

\subsubsection{The Markov Limit}

The symmetry assumptions we have made in order to obtain this $\hat{z}(t)$ are compatible
with the Markov limit in which the modes become continuous and the coupling constant
becomes flat in $k$-space (which of cause is symmetric around $\omega_0$). As in the
coherent case, the memory function $\beta{(t-s)}$ in the Markov limit equals
$\gamma\delta{(t-s)}$. Therefore in this limit the noise function is ostensibly given by
$z(t)=\rt{\gamma}\xi(t)$ where $\xi(t)$ is a real-valued Gaussian white noise term
\cite{Gar83}.

\subsection{The Linear Stochastic \sch Equation for the Quadrature Unraveling}

To find the linear non-Markovian SSE we start by applying our
assumptions to the \sch equation for the combined state
\begin{eqnarray}
   d_{t}\ket{{\Psi}(t)}&=&\Big{\{}{-i\hat{H}(t)}+\sum_{k>0} g_{k} \Big{[}
\hat{L}
 (\hat{a}_{k}\dg e^{i\Omega_{k}t}+\hat{a}_{-k}\dg e^{-i\Omega_{k}t})
  \nl- \hat{L}\dg (\hat{a}_{k} e^{-i\Omega_{k}t}+\hat{a}_{-k}
  e^{i\Omega_{k}t})\Big{]}\Big{\}} \ket{\Psi(t)}. \hspace{.5cm}
\end{eqnarray} Now by \erf{QuadratureNoiseOperatorXY}
we can write this as
\begin{eqnarray}
d_{t}\ket{ {\Psi}(t)}&=&\Big{\{}{-i\hat{H}(t)}+\hat{L}\hat{z}
  -\sum_{k>0}
g_{k}\hat{L}_{x}(\hat{a}_{k}e^{-i\Omega_{k}t}\nl+e^{i\Omega_{k}t}\hat{a}_{-k})
\Big{\}}  \ket{\Psi(t)},
\end{eqnarray} where $\hat{L}_{x}=(\hat{L}+\hat{L}\dg)$.
Using definitions  (\ref{X+}), (\ref{Y-}) and (\ref{aIintoxAndy})
we rewrite the above equation as
\begin{eqnarray} \label{TotalSchEq2}
 d_{t}\ket{{\Psi}(t)}&=& \Big{\{}{-i\hat{H}(t)}+\hat{z}\hat{L}
 -\sum_{k>0}g_{k}\hat{L}_{x} \bl\hat{X}_{k}^{+}\cos(\Omega_{k}t)
\nl
+i\hat{Y}_{k}^{+}\cos(\Omega_{k}t)-i\hat{X}_{k}^{-}\sin(\Omega_{k}t)
 \nl+\hat{Y}_{k}^{-}\sin(\Omega_{k}t)\br\Big{\}}  \ket{\Psi(t)}. \hspace{.5cm}
\end{eqnarray}

As in the coherent case to find a linear SSE we differentiate
Eq.~(\ref{LMeasurementSystemState}) with respect to time, except
that this time $\qk$ is given by Eq.~(\ref{Qstatex}) and the
ostensible probability is given by
Eq.~(\ref{QuadratureProbability}). Using Eq.~(\ref{TotalSchEq2})
we obtain \begin{widetext}\begin{eqnarray} \label{LinearNMSSE}
{\partial}_{t} \lpsiqk{t}&=&\bl {-i\hat{H}(t)}+
{z}(t)\hat{L}\br\lpsiqk{t}
 -\sum_{k>0}g_{k}\hat{L}_{x} \Big{\{}
  \cos(\Omega_{k}t)\bl i\frac{\qb
{Y}_{k}^{+}\ket{\Psi(t)}}{\rt{\Lambda(\{X_{k}^{+},Y_{k}^{-}\})}}
  +\hat{X}_{k}^{+}  \lpsiqk{t}\br
  \nl+ \sin(\Omega_{k}t)\bl\hat{Y}_{k}^{-} \lpsiqk{t}-
i\frac{\qb{X}_{k}^{-}\ket{\Psi(t)}}{\rt{\Lambda(\{X_{k}^{+},Y_{k}^{-}\})}}
\br\Big{\}}. \hspace{.8cm}
\end{eqnarray}\end{widetext}
The inner products in the above equation can be simplified to
\begin{eqnarray} \label{X-toY-}
&&\qb\hat{X}_{k}^{-}\ket{\Psi(t)}=i\frac{\partial}{\partial
Y_{k}^{-}}\qb{\Psi(t)}\rangle,  \\
\label{Y+toX+} && \qb\hat{Y}_{k}^{+}\ket{\Psi(t)}=-i\frac{\partial}{\partial
X_{k}^{+}}\qb{\Psi(t)}\rangle.
\end{eqnarray} as $\hat{X}_{k}^{\pm}$ and $\hat{Y}_{k}^{\pm}$ have the
commutators listed in
equations (\ref{com1}) and (\ref{com2}).

It can also be shown that
\begin{eqnarray}
\frac{\partial}{\partial Y_{k}^{-}}\lpsiqk{t}
&=&\frac{1}{\rt{\Lambda(\{X_{k}^{+},Y_{k}^{-}\})}}{\frac{\partial}{\partial
Y_{k}^{-}}\qb{\Psi(t)}\rangle }\nl+Y_{k}^{-}\lpsiqk{t},
 \hspace{.8cm}\\
\frac{\partial}{\partial X_{k}^{+}}\lpsiqk{t}
&=&\frac{1}{\rt{\Lambda(\{X_{k}^{+},Y_{k}^{-}\})}}{\frac{\partial}{\partial
X_{k}^{+}}\qb{\Psi(t)}\rangle
}\nl+X_{k}^{+}\lpsiqk{t},\hspace{.8cm}
\end{eqnarray}
and using equations (\ref{X-toY-}) and (\ref{Y+toX+}) with the above two
equations we can
write the inner products in terms of their conjugate variables.
This allows us to write
the linear equation as
\begin{eqnarray} \label{LinearNMSSE2}
{\partial}_{t}\lpsiqk{t} &=&\Big{\{}{-i\hat{H}(t)}+{z}(t)\hat{L}
  -\sum_{k>0}g_{k}\hat{L}_{x}
 \bl \sin(\Omega_{k}t)  \nl\times\frac{\partial}{\partial
Y_{k}^{-}} + \cos(\Omega_{k}t) \frac{\partial}{\partial
X_{k}^{+}}\br\Big{\}} \lpsiqk{t},\nl
\end{eqnarray}
which is a linear equation solely in terms of the parameters $\{X_{k}^{+}\}$
and $\{Y_{k}^{-}\}$.

As in the coherent case, to make progress towards a genuine SSE we
wish to replace the partial derivatives by a functional derivative
with respect to the noise function. To do this we note that,
\begin{eqnarray} \label{functionalX}
\frac{\partial}{\partial
X_{k}^{+}}&=&\int_{0}^{t}\frac{\delta}{\delta
z(s)}\frac{\partial z(s)}{\partial X_{k}^{+}}ds, \\
\label{functionalY} \frac{\partial}{\partial
Y_{k}^{-}}&=&\int_{0}^{t}\frac{\delta}{\delta z(s)}\frac{\partial
z(s)}{\partial Y_{k}^{-}}ds.
\end{eqnarray} Thus we  obtain
\begin{eqnarray} \label{LinearNMSSE5}
\partial_{t}\lpsizk{t}& =&\Big{\{}{-i\hat{H}(t)}+{z}(t)\hat{L}
 -\hat{L}_{x}\int_{0}^{t}\beta{(t-s)}\nl\times \frac{\delta}
  {\delta z(s)}
  ds\Big{\}}\lpsizk{t},
\end{eqnarray}
where $\beta{(t-s)}$ is the
memory function for the noise. As in the coherent state case, this
enforces an initial vacuum state for the bath. The final step to obtaining the linear
 non-Markovian SSE with real noise is to assume that the functional
derivative can be replaced by an operator as
in Eq.~(\ref{Ansatz}). With this Ansatz the linear SSE becomes
\begin{eqnarray} \label{LinearNMSSE6}
\partial_{t}\lpsizk{t} &=&\bl{-i\hat{H}(t)}+{z}(t)\hat{L}
  -\hat{L}_{x}\int_{0}^{t}\beta(t-s)
  \nl\times\hat{O}_{z}(t,s)
  ds\br\lpsizk{t}.
\end{eqnarray}

\subsubsection{The Markov Limit}

Finally in this subsection we determine the Markov limit of
this equation. Applying the results at the end of Sec.~\ref{quadOperatorSection}, we
get
\begin{equation} \label{LinearMSSE2}
\partial_{t}\lpsizk{t} =\bl{-i\hat{H}(t)}+ \hat{L}z(t)
  -\frac{\gamma}{2}\hat{L}_{x}\hat{L}\br\lpsizk{t},
\end{equation} as $\hat{O}_z(t,t)=\hat{L}$.
This is in \str from. We transform this to the \ito form
by using the method in Sec.~\ref{sec:itostr1}. In this case
 the \ito correction is
\begin{equation} \frac{d t}{2}\sum_{l}\bl b_j
\frac{\partial}{\partial\psi_l}b_j+ b_j^{*}
\frac{\partial}{\partial\psi_l^*}b_j\br =\frac{\gamma
dt}{2}\sum_{l,k,}L_{j,l}L_{l,k}\psi_{k},
\end{equation} and the \ito SSE is
\begin{equation} \label{LinearMSSE3}
d\lpsizk{t}=dt\bl\frac{-i\hat{H}(t)}{\hbar}+\hat{L}z(t)
  -\frac{dt\gamma}{2}\hat{L}\dg\hat{L}\br\lpsizk{t},
\end{equation} which is the general linear homodyne SSE \cite{GoeGra94,Wis96}
as $z(t)=\rt\gamma\xi(t)$.
\vspace{3ex}

\subsection{The Actual Stochastic \sch Equation for the Quadrature Unraveling}

As in the coherent case, to find an actual SSE (generating states
with the actual probability) we need to find  random variables
with the actual probabilities of measurement results $\{q_{k}\}$.
To sort these out we use the Girsanov transform (\ref{Girsanov})
to find a first-order partial differential equation (PDE) for the
probability, from which the characteristic equation generates the
transformed variables
\begin{eqnarray}
  \partial_{t}{ P}(\{X_{k}^{+},Y_{k}^{-}\},t)&=&\bl\lpsiqb{t}
 { {\partial}_{t}}\lpsiqk{t} +{\rm c.c}
\br\nl\times\Lambda(\{X_{k}^{+},Y_{k}^{-}\}).
\end{eqnarray}
Using equations (\ref{LinearNMSSE2}) allows us to write
\begin{eqnarray}
\partial_{t}{ P}(\{X_{k}^{+},Y_{k}^{-}\},t)&=&-\sum_{k>0}
g_{k}\frac{\partial}{\partial X_{k}^{+}}\bl\cos(\Omega_{k}
t)\lpsiqb{t}\nl\times\hat{L}_x \lpsiqk{t}
\Lambda(\{X_{k}^{+},Y_{k}^{-}\})\br\nl -\sum_{k>0}
g_{k}\frac{\partial}{\partial Y_{k}^{-}}\bl\sin(\Omega_{k}
t)\lpsiqb{t} \nl\times\hat{L}_x \lpsiqk{t}
\Lambda(\{X_{k}^{+},Y_{k}^{-}\})\br.\nl
\end{eqnarray}
This can be simplified to
\begin{eqnarray}
&&\partial_{t}{ P}(\{X_{k}^{+},Y_{k}^{-}\},t)=\nl-\sum_{k>0}
g_{k}\frac{\partial} {\partial X_{k}^{+}}\bl\cos(\Omega_{k}
t)\langle \hat{L}_{x}\rangle_t { P}(\{X_{k}^{+},Y_{k}^{-}\},t)\br
\nl-\sum_{k>0} g_{k}\frac{\partial}{\partial
Y_{k}^{-}}\bl\sin(\Omega_{k} t)\langle \hat{L}_{x}\rangle_t {
P}(\{X_{k}^{+},Y_{k}^{-}\},t)\br.\nl
\end{eqnarray}
where $\langle \hat{L}_{x}\rangle_t$ is defined by equation
(\ref{Expect}).

The characteristic equations are
\begin{eqnarray} \label{X+Derivative}
\frac{d  }{dt}X_{k}^{+}&=&g_{k}\cos(\Omega_{k} t)\langle \hat{L}_{x}\rangle_t ,\\
\label{Y-Derivative} \frac{d }{dt}Y_{k}^{-}&=&g_{k}\sin(\Omega_{k}
t)\langle \hat{L}_{x} \rangle_t.
\end{eqnarray}
Integrating these differential equation from time $0$ to $t$ we get
\begin{eqnarray}
 X_{k}^{+}(t) &=&X_{k}^{+}(0)+\int_{0}^{t}g_{k}\cos(\Omega_{k} s)\langle
\hat{L}_{x}\rangle_s ds,\\
Y_{k}^{-}(t)&=&Y_{k}^{-}(0)+\int_{0}^{t}g_{k}\sin(\Omega_{k}
s)\langle \hat{L}_{x} \rangle_s ds.
\end{eqnarray}
The distribution for $X_{k}^{+}(0)$ and $Y_{k}^{-}(0)$ is due to
the quantum initial conditions. As before, the use of the
functional derivative in \erf{LinearNMSSE5} implies that the
initial bath state is a vacuum state. Thus, the randomness in
$X_{k}^{+}(0)$ and $Y_{k}^{-}(0)$ is that of the ostensible
distribution:
\begin{equation}
{ P}(\{X_{k}^{+},Y_{k}^{-}\},0)= \Lambda(\{X_{k}^{+},Y_{k}^{-}\})=\frac{e^{-\sum_{k>0}
({X_{k}^{+}}^2+{Y_{k}^{-}}^2)}}{\pi^{\kappa/2}}.
\end{equation}

With the above random variable equations for $X_{k}^{+}(t)$ and
$Y_{k}^{-}(t)$ we can write the noise function for the actual
probability as
\begin{eqnarray} \label{TrueNoise}
{z}(t)&=& z_{\Lambda}(t)+\int_{0}^{t}\langle\hat{L}_{x}\rangle_t \beta(t-s) ds,
\end{eqnarray} where $z_{\Lambda}(t)$ is the random variable
with statistics determined by the
$\Lambda(\{X_{k}^{+},Y_{k}^{-}\})$ distribution. That is, the
correlations of $z_{\Lambda}(t)$ are those of $z(t)$ in
 Eq.~(\ref{NMNoiseCorrelation}).

Now we have the correct noise function we can calculate the actual
SSE. As in the coherent case  we need $\partial_t\lpsizk{t}$,
and for this case equation
(\ref{diffTerm1}) will be
\begin{eqnarray}
{\partial_t}\lpsizk{t} &=&\Big{\{}-{i \hat{H}(t)}+ \hat{L}z(t)-
{(\hat{L}_{x}-\langle\hat{L}_{x}\rangle_t)}\nl\times\int_{0}^{t}
\beta(t-s)\frac{\delta}{\delta {z}(s)} ds \lpsizk{t}.
\end{eqnarray} Following the same procedure as in the coherent case we obtain
\begin{widetext}\begin{eqnarray}
\partial_{t}\psizk{t}&=& \bl-{i
\hat{H}(t)}+(\hat{L}-\langle\hat{L}\rangle_t){z}(t)\br\psizk{t}
 -\frac{1}{|\tilde{\psi}_{\{q_k\}}(t)|}(\hat{L}_{x}-\langle\hat{L}_{x}\rangle_t)
\int_{0}^{t} \beta(t-s)\frac{\delta}{\delta{z}(s)} ds \lpsizk{t} \nl +
\frac{1}{|\tilde{\psi}_{\{q_k\}}(t)|}\langle(\hat{L}_{x}-
\langle\hat{L}_{x}\rangle_t)\int_{0}^{t}
\beta(t-s)\frac{\delta}{\delta {z}(s)} ds \lpsizk{t} \psizk{t}.
\end{eqnarray}\end{widetext}
Again this is not a  SSE until we make the Ansatz
defined in Eq.~(\ref{Ansatz}), which gives
\begin{eqnarray} \label{NormalisedNMSSEFinal}
\partial_{t}\psizk{t}&=& \bl-{i
\hat{H}(t)}+(\hat{L}-\langle\hat{L}\rangle_t){z}(t)\nl
-(\hat{L}_{x}-\langle\hat{L}_{x}\rangle_t)\int_{0}^{t}
\beta(t-s)\hat{O}_z(t,s) ds \nl +
\Big{\langle}(\hat{L}_{x}-\langle\hat{L}_{x}\rangle_t)\int_{0}^{t}
\beta(t-s)\hat{O}_z(t,s) ds\nl\times\Big{\rangle}_t\br \psizk{t}.
\end{eqnarray}
This is the actual SSE for real-valued noise. All of the comments
regarding the interpretation of the corresponding complex-valued
noise SSE (\ref{CSSEOperatorForm}) carry over to this case.

\subsubsection{The Markov Limit}

Taking the Markov limit of the actual SSE results
in a noise function of the form
\begin{equation}\label{MnoiseFunctionQuad}
 {z}(t)=z^{\Lambda}(t)+\frac{\gamma}{2}\langle\hat{L}_{x}\rangle_t,
\end{equation}
where $z_{\phi}^{\Lambda}(t)= \rt{\gamma}\xi(t)$.
The actual SSE becomes
\begin{eqnarray}
\partial_{t}\psizk{t}&=& \bl-{i \hat{H}(t)}+ (\hat{L}-\langle\hat{L}\rangle_t)
({z}(t)+\frac{\gamma}{2}\langle\hat{L}_{x}\rangle_t)\nl-
\frac{\gamma}{2}(\hat{L}_{x}\hat{L}-\langle\hat{L}_{x}\hat{L}\rangle_t)\br\psizk
{t},
\end{eqnarray}
This is in Stratonovich form, to compare it to the equivalent homodyne SSE we need to
convert it to \ito form. The \ito correction term for this equation is
\begin{eqnarray}
&&\frac{dt}{2}\sum_{l} \bl b_l \frac{\partial}{\partial \psi_l}
b_j+b^{*}_{l}\frac{\partial}{\partial \psi^{*}_{l}} b_j \br =
\frac{dt\gamma}{2}\bl \hat{L}\hat{L}
-2\hat{L}\langle\hat{L}\rangle_t\nl\hspace{1cm}-\langle\hat{L}_{x}\hat{L}\rangle_t
+\langle\hat{L}_{x}\rangle_t\langle\hat{L}\rangle_t
+\langle\hat{L}\rangle_t\langle\hat{L}\rangle_t\br\psizk{t}.\nl
\end{eqnarray} Adding this to the Stratonovich SSE we get the following
\ito SSE,
\begin{eqnarray} \label{NormalisedMSSEito}
d\psizk{t}&=&dt\Big{\{}-{i
\hat{H}(t)}+(\hat{L}-\langle\hat{L}\rangle_t
)z(t)-\frac{\gamma}{2}dt\bl\hat{L}\dg\hat{L}\nl-\hat{L}\langle\hat{L}\dg\rangle_
t +\hat{L}\langle\hat{L}\rangle_t
-\langle\hat{L}\rangle_t\langle\hat{L}\rangle_t\br \Big{\}}
\psizk{t}.\nl
\end{eqnarray}
 This is the same as the homodyne SSE presented in Ref. \cite{WisMil93c,DorNie00}
 when we substitute in \erf{MnoiseFunctionQuad} for $z(t)$. As in the coherent case
 there will be a difference between $z(t)$ and the homodyne current, which from reference
 \cite{WisMil93c} is $I(t)=\rt{\gamma}\xi(t)+{\gamma}\langle\hat{L}_{x}\rangle_t$. This
 difference again comes down to the fact the in the quantum trajectory theory the
 measurement occurs a time $dt$ later.

\section{A Simple System}

In this section we apply the above theory to a very simple non-Markovian system: a TLA
coupled linearly and with the same strength to two single mode fields (labeled by $k=\pm
1$) that are detuned from $\omega_0$ by $\pm \Delta$ respectively. Without loss of
generality, we can take the coupling strength $g_{1}=g$ to be real. Then the memory
function becomes
\begin{equation}\label{MemoryFunctionSimpleSys}
  \alpha{(t-s)}=2 g^2 \cos(\Delta{(t-s)}).
\end{equation}
Note that this memory never decays, indicating that the dynamics of the atom is extremely
non-Markovian. This is different from all cases considered by DSG, where the memory was
taken to decay exponentially. It is thus interesting to see how the formalism copes with
this extreme case. At the same time, the simplicity of the bath (two modes) means that an
exact numerical solution for $\rho_{\rm red}(t)$ is relatively easy to find. This allows
verification of the validity of the SSEs in reproducing $\rho_{\rm red}(t)$ by ensemble
average, for both the linear and actual (nonlinear) cases.

We would also like to see the
different individual behaviour of the trajectories corresponding to
two different measurements
(coherent state and quadrature measurements). This is readily apparent
in this system for the initial condition
$\ket{\psi(0)}=\ket{e}$, where
$\ket{e}$ and $\ket{b}$ are the excited and ground state of the TLA,
respectively, so we choose this for all our simulations.


\subsection{Exact Solution}
\begin{figure}
\includegraphics[width= .45\textwidth]{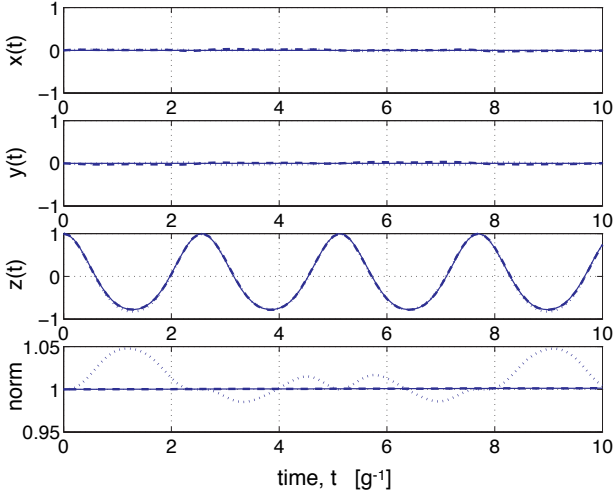}
\caption{\label{CoherentAve} This figure depicts the reduced state
calculated by three different methods; the exact solution (solid
line), the ensemble average of 1000 SSEs for both the linear
(dotted) and actual (dashed) SSE for the coherent unraveling. In
this figure all calculations were done using a simple Euler method
with
 a step size of $dt=0.0001$, a detuning of
$\Delta=2g$ and initial system state of the form $\ket{\psi(0)}=\ket{e}$.}
\end{figure}

 To calculate the exact
$\rho_{\rm red}(t)$ we need to solve the \sch equation, which is
displayed in Eq.~(\ref{IntSchEquation}). For this simple
system we assume $\hat{H}=0$ and
\begin{equation} \label{IntHamSimpleSys}
 \hat{V}(t)=g e^{i\Delta
 t}(\hat{a}\dg_{1}\hat{\sigma}-\hat{a}_{-1}\hat{\sigma}\dg)+g
  e^{-i\Delta t}(\hat{a}\dg_{-1}\hat{\sigma}-\hat{a}_{1}\hat{\sigma}\dg)
\end{equation}
as $\Omega_{1}=\Delta=-\Omega_{-1}$ and $g=g_{-1}=g_{1}$. Here
the Lindblad operator $\hat{L} = \hat{\sigma} = \ket{b}\bra{e}$.
Since initially the field is in the vacuum state
($\ket{0_{1}}\otimes\ket{0_{-1}}$) then the only non-zero
complex amplitudes in
$\ket{\Psi(t)}$ are
\begin{equation}
\ket{\Psi(t)}=c_{1}(t)\ket{b00}+c_2(t)\ket{e00}+c_3(t)\ket{b01}+c_4(t)\ket{b10},
\end{equation} where $\ket{b00}$ is short hand for
$\ket{b}\otimes\ket{0_{1}}\otimes\ket{0_{-1}}$ {\em etc.}
 Applying the above Hamiltonian to this state
  we get the following four differential equations for the
complex amplitudes,
\begin{eqnarray}
 \dot{c}_1(t)&=&0 ,\\
  \dot{c}_2(t)&=&-c_3(t)ge^{i\Delta t}-c_4(t) g e^{-i\Delta t} ,\\
   \dot{c}_3(t)&=&c_2(t) g e^{-i\Delta t},\\
    \dot{c}_4(t)&=&c_2(t) g e^{i\Delta t},
\end{eqnarray} which can be solved numerically. For
 the initial state $\ket{e00}$, $c_2(0)=1$ and the rest
 are zero. Once we have the amplitudes for all time we know $\ket{\Psi(t)}$
and by Eq.~(\ref{ReducedState}) we can then calculate $\rho_{\rm red}(t)$.
For the
TLA it is convenient to define the
reduced state in terms of a pseudo-spin vector $(x,y,z)$ by
\begin{equation}
  \rho_{red}(t)=\frac{1}{2}[I+x(t)\sigma_{x}+y(t)\sigma_{y}+z(t)\sigma_{z}],
\end{equation}
where $x(t)$, $y(t)$ and $z(t)$ are real parameters which equal the expected
 value of the corresponding spin matrix. These can be found from the
 above complex amplitudes by
\begin{eqnarray}
I&=& |c_1(t)|^2+|c_2(t)|^2+|c_3(t)|^2+|c_4(t)|^2,\\
x(t)&=& c_2(t)c_1^*(t)+c_2^*(t)c_1(t),\\
y(t)&=& -ic_2(t)c_1^*(t)+ic_2^*(t)c_1(t),\\
z(t)&=&|c_2(t)|^2-|c_1(t)|^2-|c_3(t)|^2-|c_4(t)|^2.
\end{eqnarray}
To  graphically illustrate the reduced state we numerically calculated the above real
parameters for $\Delta=2g$. The results are shown in Fig. \ref{CoherentAve} as a solid
line.

\subsection{Coherent Unraveling}

For the simple system the memory function, Eq.~(\ref{MemoryFunction})
is given by \erf{MemoryFunctionSimpleSys}, and
and the noise operator for the coherent unraveling is,
\begin{equation}
 \hat{z}(t)=g\hat{a}_{1}e^{-i\Delta{t}}+g\hat{a}_{-1}e^{i\Delta{t}}.
\end{equation}
The linear SSE was obtained when we assumed an ostensible probability
 $\Lambda(a_1,a_{-1})$ equal to the vacuum distribution
\begin{equation}\label{SimpleSysProb}
\Lambda(a_1,a_{-1})=\pi^{-2}e^{-|a_1|^2-|a_{-1}|^2} .
\end{equation}
With this probability distribution, we can write
the noise function as a random variable
equation of the form,
\begin{equation}
{z}(t)=g{a}_{1}e^{-i\Delta{t}}+g{a}_{-1}e^{i\Delta{t}},
\end{equation}
where $a_1$ and $a_{-1}$ are complex GRVs of mean 0 and variance 1.

Applying the simple systems dynamics to equation
(\ref{CLSSEFunctionalForm}) we obtain
\begin{equation}\label{CLSSEsimpleSys}
  \partial_{t} \lpsizk{t}=\Big{(}z^{*}(t)\hat{\sigma}
  -\hat{\sigma}\dg\int_{0}^{t}\alpha{(t-s)}\frac{\delta}{\delta z^{*}(s)}
ds\Big{)}\lpsizk{t}.
\end{equation}
 In Sec.~\ref{CoherentLSSE}
 we made the general Ansatz described by Eq.~(\ref{Ansatz}).
For this simple system the specific Ansatz we will use is
\begin{equation} \label{SimpleAnsatz}
\frac{\delta}{\delta
z^{*}(s)}\lpsizk{t}=f(t,s)\hat\sigma\lpsizk{t}.
\end{equation} To work out the functions $f(t,s)$ we use the
following consistency condition \cite{DioGisStr98},
\begin{equation}\label{ConsistyencyCondition}
\frac{\delta}{\delta
z^{*}(s)}\frac{\partial}{\partial
t}\lpsizk{t}=\frac{\partial}{\partial t}\frac{\delta}{\delta
z^{*}(s)}\lpsizk{t},
\end{equation}
This gives
\begin{equation}\label{fSimpleSys}
  \partial_t f(t,s)\hat\sigma \lpsizk{t}=f(t,s)F(t)\hat\sigma\lpsizk{t},
  \end{equation} where
\begin{equation} \label{FSimpleSys}
F(t)=\int_{0}^{t}\alpha{(t-s)}f(t,s)ds.
\end{equation}
This allows us to write the linear SSE for the coherent unraveling as
\begin{equation}\label{CLSSEsimpleSys2}
  \partial_{t} \lpsizk{t}=\Big{[}z^{*}(t)\hat{\sigma}
  -\hat{\sigma}\dg\hat{\sigma}F(t)\Big{]}\lpsizk{t}.
\end{equation} This is simple to to solve numerically provide we have a solution for
$F(t)$.

The best way to calculate $F(t)$ is to split it into to two terms,
$F(t)=F_1(t)+F_{-1}(t)$
where,
\begin{equation}
F_{1}(t)=\int_{0}^{t}|g|^2e^{-i\Delta(t-s)}f(t,s)ds=F_{-1}^{*}(t).
\end{equation}
Differentiating the above equations for $F_1(t)$ and $F_{-1}(t)$
and using Eq.~(\ref{fSimpleSys}) and the fact that $f(t,t)=1$ yields
\begin{eqnarray} \label{Fa}
d_t F_{1}(t)&=&|g|^2-i\Delta F_{1}(t)+F_{1}(t)F(t), \\ \label{Fb} d_t
F_{-1}(t)&=&|g|^2+i\Delta F_{-1}(t)+F_{-1}(t)F(t),
\end{eqnarray} which can be solved numerically. The initial conditions are
 $F(0)=F_{1}(0)=F_{-1}(0)=0$.
Writing $\lpsizk{t}=C_e(t)\ket{e}+C_{b}(t)\ket{b}$ gives us the following two differential
equations,
\begin{eqnarray}
d_t C_{e}(t)&=&-C_e(t)F(t), \\
d_t C_{b}(t)&=&z^{*}(t)C_e(t).
\end{eqnarray} For an excited-state initial condition
these equation can be solve numerically. Note that these solutions
will not remain normalized, and the norm of most of them becomes very
small. This reflects the fact that a typical
individual solution of this SSE
does not correspond to a typical measurement result.
Nevertheless,  the ensemble average of the unnormalized states
is $\rho_{\rm red}(t)$. To show this we simulated 1000 SSE for
different $z(t)$. The results of this simulation are shown in Fig.
\ref{CoherentAve} as a dotted line, where the agreement with the exact
solution is good.

The actual SSE for coherent unraveling is found by applying the
above results to Eq.~(\ref{CSSEOperatorForm}). Doing this we obtain
\begin{eqnarray}
  \partial_t
\psizk{t}&=&\Big{\{}(\hat\sigma-\langle\hat{\sigma}\rangle_{t}) z^{*}(t)
  -(\hat\sigma\dg-\langle\hat{\sigma}\dg\rangle_{t})\hat\sigma F(t)
  \nl+\Big{\langle}(\hat\sigma\dg-\langle\hat{\sigma}\dg\rangle_{t})
  \hat\sigma \Big{\rangle}_{t} F(t)
  \Big{\}}\psizk{t}.
\end{eqnarray}
The noise, $z^{*}(t)$ in this equation is given by,
\begin{equation}
  z^{*}(t)=z_{\Lambda}^{*}(t)+\int_{0}^{t}\alpha^{*}{(t-s)}\langle
\hat\sigma\dg\rangle_s ds,
\end{equation}
where $z_{\Lambda}(t)$ is the noise function used in the linear case.
With this SSE the
two differential equations for the complex amplitudes become
\begin{eqnarray}
 d_t C_{e}(t)&=&-C_{e}^{2}(t)C_b^{*}(t)z^{*}(t)+F(t)C_e(t)(-1\nl+|C_e(t)|^2-
 |C_e(t)|^2|C_b(t)|^2), \\
d_t
C_{b}(t)&=&C_e(t)(1-|C_b(t)|^2)z^{*}(t)+F(t)C_b(t)|C_e(t)|^2\nl\times(2-|C_b(t)|
^2).
\end{eqnarray}

\begin{figure}
\includegraphics[width= .45\textwidth]{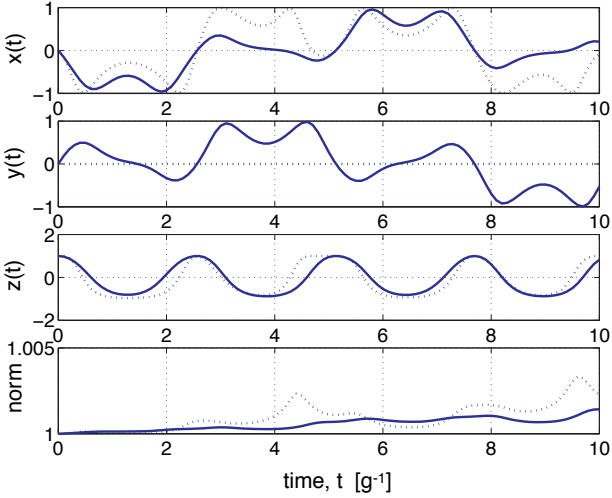}
\caption{\label{Trajectories} This figure shows a typical trajectory
generated by the actual SSE for both the
coherent (solid) and quadrature (dotted) unraveling. These were  all done with the
parameters defined in Fig. \ref{CoherentAve}.}
\end{figure}

The solution to these equations is an actual state, in the sense that it is normalized,
and generated with the actual probabilities. Thus a typical trajectory does give, at any
time $t$, a typical state that corresponds to an observer measuring it at that time in the
coherent basis. It is thus worth examining a typical trajectory, which we have plotted in
Fig. \ref{Trajectories} (the solid line). The normalization of the state is shown to
remain equal to one, within the error introduced by the integration algorithm.  To show
that the ensemble average of these trajectories is the reduced state, an ensemble average
of 1000 SSE was simulated and the results are depicted in Fig. \ref{CoherentAve} (dashed
line). We see that the actual case is closer to the $\rho_{\rm red}(t)$ then the linear
case. This is expected as in general the linear SSE converges slower than the actual SSE,
as most of the states generated from the linear SSE have virtually no contribution to the
mean.

\subsection{Quadrature Unraveling}

If we apply the theory for the quadrature unraveling to this
simple system, the quadrature noise operator, equation
(\ref{QuadratureNoiseOperatorXY}) becomes
\begin{equation}
 \hat{z}(t)=2g\{\hat{X}_{1}^{+}\cos(\Delta{t)})+\hat{Y}_{1}^{-}\sin(\Delta{t
})\},
\end{equation} and the quadrature noise function is
\begin{equation}
 {z}(t)=2g\{{X}_{1}^{+}\cos(\Delta{t})+{Y}_{1}^{-}\sin(\Delta{t})\},
\end{equation} which is real.
If we choose the ostensible probability to equal the vacuum probability, then
\begin{equation}\label{SimpleSysProb2}
\Lambda(X_1^{+},Y_{1}^{-})=\pi^{-1}{e^{-{X_1^{+}}^2-{Y_1^{-}}^2}}.
\end{equation}
Thus for the linear case ${X}_{1}^{+}$ and ${Y}_{1}^{-}$ are
GRVs of mean zero and variance $1/2$.

For this simple system the quadrature linear SSE, Eq.~(\ref{LinearNMSSE5}),
becomes
\begin{equation}\label{QLSSEsimpleSys}
  \partial_{t} \lpsizk{t}=\Big{(}z(t)\hat{\sigma}
  -\hat{\sigma}_{x}\int_{0}^{t}\beta{(t-s)}\frac{\delta}{\delta z(s)}
ds\Big{)}\lpsizk{t}
\end{equation} As
for the coherent case we can make an Ansatz for the functional
derivative. We again choose Eq.~(\ref{SimpleAnsatz}).
 This allows us to write the
quadrature linear SSE as
\begin{equation}\label{QLSSEsimpleSys2}
  \partial_{t} \lpsizk{t}=\Big{(}z(t)\hat{\sigma}
  -\hat{\sigma}_{x}\hat{\sigma}F(t)\Big{)}\lpsizk{t},
\end{equation} where
$F(t)$ is given by
\begin{equation} \label{FSimpleSys2}
F(t)=\int_{0}^{t}\beta{(t-s)}f(t,s)ds,
\end{equation}
 and $\beta{(t-s)}=2|g|^{2}\cos(\Delta(t-s))$.

\begin{figure}
\includegraphics[width= .45\textwidth]{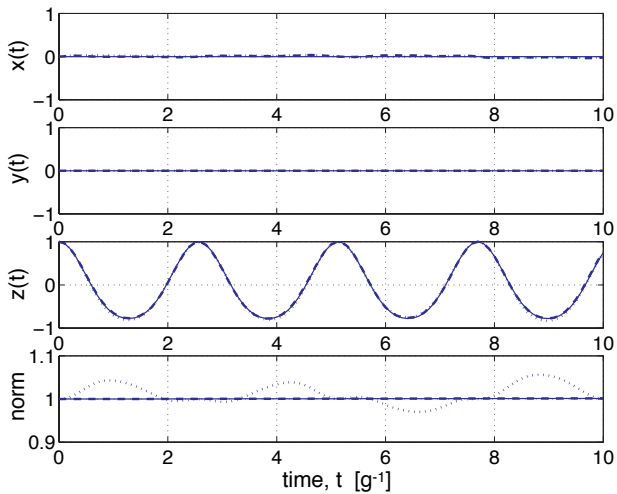}
\caption{\label{QuadratureAve} This figure depicts the reduced
state calculated by three different methods; the exact solution
(solid line), the ensemble average of 1000 SSEs for both the
linear (dotted) and actual (dashed) SSE for the quadrature
unraveling. These where all done with the parameters defined in
Fig. \ref{CoherentAve}.}
\end{figure}

It turns out for this simple system $F(t)$ is the same for both
the coherent and quadrature unraveling, because
 $\alpha{(t-s)}=\beta{(t-s)}$. Knowing $F(t)$, we get the following two
differential equations for the state:
\begin{eqnarray}
d_t C_{e}(t)&=&-C_e(t)F(t), \\
d_t C_{b}(t)&=&z(t)C_e(t).
\end{eqnarray}
These are the same as for the coherent case, except that $z(t)$ is generated differently.
To show that the ensemble average of the solution to the linear SSE for the quadrature
unraveling converges to $\rho_{\rm red}(t)$, 1000 trajectories for different $z(t)$ where
simulated. The results of these simulations are shown in Fig. \ref{QuadratureAve} as a
dotted line, where it is seen that the ensemble average of the linear SSE does reproduce
the exact solution for $\rho_{\rm red}(t)$ with little error.

The actual SSE for quadrature unraveling is found by applying the above results to
Eq.~(\ref{NormalisedNMSSEFinal}),
\begin{eqnarray}
  \partial_t
\psizk{t}&=&\Big{\{}(\hat\sigma-\langle\hat{\sigma}\rangle_{t})
z(t)
  -(\hat\sigma_{x}-\langle\hat{\sigma}_{x}\rangle_{t})\hat\sigma F(t)
  \nl+\Big{\langle}(\hat\sigma_{x}-\langle\hat{\sigma}_{x}\rangle_{t})\hat\sigma
  \Big{\rangle}_{t} F(t)   \Big{\}}\psizk{t}.
\end{eqnarray}
The noise, $z(t)$ in this equation is given by,
\begin{equation}
  z(t)=z_{\Lambda}(t)+\int_{0}^{t}\beta{(t-s)}\langle
\hat\sigma_{x}\rangle_s ds,
\end{equation}
where $z_{\Lambda}(t)$ is the noise function used in the linear case. With
this SSE the
two differential equations for the complex amplitudes become,
\begin{eqnarray}
 d_t C_{e}(t)&=&F(t)C_e(t)(-1+|C_e(t)|^2-
 |C_e(t)|^2|C_b(t)|^2)\nl-F(t)C_e^{3}(t){C_b^{*}}^{2}(t)-C_{e}^{2}(t)C_b^{*}(t)z(t) ,\\
d_t C_{b}(t)&=&F(t)C_b(t)|C_e(t)|^2 (2-|C_b(t)|^2)
\nl+F(t)C_b^{*}(t)
C_e^2(t)(1-|C_b(t)|^2)+C_e(t)\nl\times(1-|C_b(t)|^2)z(t).
\end{eqnarray}

A typical trajectory from the quadrature SSE is illustrated in
Fig. \ref{Trajectories} (the dotted line). Note the feature that
clearly distinguishes it from the coherent trajectory: $y$ is always
zero. To show that the
solution of the actual SSE reproduces the reduced state on average, an
ensemble of 1000 actual SSEs was simulated and the results are
depicted in Fig. \ref{QuadratureAve} (dashed line). We see that it
reproduces the exact solution, again  with less error than that from
the linear SSE.

\section{Discussion and Conclusions}

In this paper we have explored non-Markovian stochastic \sch equations by furthering the
work of Diosi, Strunz and Gisin \cite{Str96,Dio96,DioStr97,DioGisStr98,StrDioGis99}.
Specifically, we have interpreted  their results in the framework of quantum measurement
theory. Their SSEs arise as a special case when the measurement basis of the bath is the
coherent states, so we label it the coherent unraveling.  The benefit of using the
measurement interpretation is two-fold.

First, it allows us a better understanding of the interpretation of
non-Markovian SSEs. The state at any time $t$ generated by the
SSE can be interpreted as a conditioned system state, given a
particular result from a particular measurement on the bath.
However, the measurements at different times are incompatible, so the
linking together of different states over time is, we have argued, a
convenient fiction.  Thus the trajectory generated by a non-Markovian
SSE does not have the same physical status as that generated by a
Markovian SSE, where the measurements at different times are
compatible and the states at different times can represent a single
evolving system.

Second, it allows us to generate other sorts of SSEs corresponding to different sorts of
measurements on the bath (unravelings). In this paper we presented a second unraveling,
based on measuring certain quadrature operators on the bath. This gives rise to an SSE
only under certain assumptions to do with the bath frequencies and couplings. The
resultant SSE contains real-valued noise, as opposed to the complex noise in the SSE of
DSG. The ability to construct a non-Markovian SSE with real-valued noise is contrary to
the expectation expressed by DSG in \cite{DioGisStr98}.

We have also shown in this paper that the Markov limit of the
quadrature and coherent unravelings are homodyne and heterodyne
detection respectively. As noted above,
in this Markov limit the SSE generates a true
quantum trajectory for a conditioned system state over time. It is
interesting that this arises smoothly as the limit of a non-Markovian
SSE that does not have this interpretation. However, as we have shown,
one has to be very careful with the definition of the time of
measurement in order to reconcile this limit with the usual quantum
trajectory theory.

To illustrate our general
theory we have applied it to a simple system: a TLA coupled linearly to
just two
single-mode fields detuned from the atom by $\pm \Delta$.
This is an extremely non-Markovian problem with no finite memory time,
unlike the previous examples considered by DSG. Nevertheless the
theory is able to describe the evolution of the atom by an SSE. In
Fig.~\ref{Trajectories} we displayed typical non-Markovian SSEs for both the
quadrature and coherent unraveling, and in Figs.~\ref{CoherentAve} and
\ref{QuadratureAve} we showed that on average both SSEs do generate
the exact reduced state.

In conclusion this paper has presented a significant generalization of the DSG approach to
non-Markovian SSE.  However, there is still a lot of questions to be answered.

First, is it possible within this framework to derive other classes
of non-Markovian SSEs? In particular, is it possible to describe
an unraveling based on discrete measurement on the bath, say the in
number-state basis?

Second, is there a physical system where our theory could be
naturally applied? That is, is there a physical system where the bath
could be measured in a suitable basis at an arbitrary time so as to
produce a pure conditioned system state?

Third, what conditions are necessary for one to be able to find a
suitable Ansatz for replacing the functional derivative with an
operator? As we have argued, this is necessary to create a genuine
SSE. Yu, Di\'osi, Gisin and Strunz have given a general procedure for
finding this operator, but only when the system dynamics are weakly
non-Markovian (the so-called `post-Markovian' approximation)
 \cite{YuDioGisStr99,YuDioGisStr00}. We suspect that the conditions
 for finding an exact Ansatz depend upon both the nature of the
 system and its coupling to the bath.

Fourth, can the techniques of non-Markovian SSEs be applied as a
numerical tool for studying
real systems? We have in mind potentially strongly non-Markovian systems
 such as an atom laser \cite{Hop}
or photon emission in a photonic band-gap material
\cite{Joh84,BayLamMol97}.

Fifth, and last, is there an alternative framework to standard quantum
measurement
theory in which there is a physical interpretation for a trajectory generated
by a non-Markovian SSE? That is, can the
states at different times in a single trajectory
generated by the SSE be interpreted as pertaining to a single system
in some non-standard approach to quantum measurements? This is a very
open question.

\end{document}